\documentclass [aps,prb,showpacs,twocolumn,amsmath,amssymb,superscriptaddress]{revtex4-1}

\usepackage{subfigure}
\usepackage{amsmath}    

\usepackage{graphicx}   
\usepackage{subfigure}  

\usepackage{color}

\begin{document}

\title{Electronic, vibrational and transport properties of pnictogen substituted ternary skutterudites}
\author{Dmitri Volja}
\affiliation{Department of  Materials Science and
Engineering,  Massachusetts Institute of Technology,  Cambridge,
MA 02139}
 \email{dvolja@mit.edu}   

\author{Boris Kozinsky}
\affiliation{Research and Technology Center, Robert Bosch LLC, Cambridge, MA 02142, U.S.A.
}

\author{An Li}
\affiliation{Research and Technology Center, Robert Bosch LLC, Cambridge, MA 02142, U.S.A.
}
\affiliation{Department of  Materials Science and
Engineering,  Massachusetts Institute of Technology,  Cambridge,
MA 02139}

\author{Daehyun Wee}
\altaffiliation{Research and Technology Center, Robert Bosch LLC, Palo Alto, California 94304, USA}

\affiliation{
Department of Environmental Science and Engineering,
Ewha Womans University,  Seoul, 120-750, Korea }

\author{Nicola Marzari}

\affiliation{Theory and Simulation of Materials, \'Ecole Polytechnique F\'ed\'eral de Lausanne, 1005 Lausanne, Switzerland}
\affiliation{Department of  Materials Science and
Engineering,  Massachusetts Institute of Technology,  Cambridge,
MA 02139}

\author{Marco Fornari}
\affiliation{Department of Physics,  Central Michigan University, Mount Pleasant,
MI 48859}

\begin{abstract}

First principles calculations are used to investigate electronic band structure and vibrational spectra of pnictogen substituted ternary skutterudites. We compare the results with the prototypical binary composition CoSb$_3$ to identify the effects of substitutions on the Sb site, and evaluate the potential of ternary skutterudites for thermoelectric applications. Electronic transport coefficients are computed within the Boltzmann transport formalism assuming a constant relaxation time, using a new methodology based on maximally localized Wannier function interpolation. Our results point to a large sensitivity of the electronic transport coefficients to carrier concentration and to scattering mechanisms associated with the enhanced polarity. The ionic character of the bonds is used to explain the detrimental effect on the thermoelectric properties.
\end{abstract}

\pacs{72.20.-i,63.20.D-,71.15.Mb}
\date{\today}

\maketitle

\section{Introduction}
Thermoelectric materials with the filled skutterudite structure are considered to be a
prototypical realization of the phonon-glass electron-crystal paradigm (PGEC) proposed
by Slack.\cite{rowe_g._1995,cahill_lower_1992} Indeed many compositions in this structural family  exhibit low thermal
conductivities ($k \simeq 0.5-5 $ Wm$^{-1}$K$^{-1}$), Seebeck coefficients (S) from -200 $\mu V/K$ to 200 $\mu V/K$, and
electrical resistivities ($\rho$) in  the range from 10$^{-4}$ to 10$^{-3}$  $\Omega \cdot cm$
at room temperature, depending on doping levels.\cite{fleurial_skutterudites:_1994} Their figure of merit ZT  ($ZT =T S^2/\rho k$
is used to characterize the material's performance)~\cite{snyder_complex_2008} reaches values in excess of 1.4 at high temperature
in the bulk form.\cite{fleurial_skutterudites:_1997,vineis_nanostructured_2010}
Skutterudites have been investigated for use in high-reliability thermoelectric modules designed for space applications,
\cite{fleurial_skutterudites:_1997} owing to their good thermal stability and mechanical strength throughout the operating
temperature range. Mechanical strength is of particular importance in automotive and household applications,~\cite{sakamoto_skutterudite-based_2011}
 where stress due to repeated thermal cycling is a major engineering challenge. The chemical robustness and stability of the
 skutterudite crystal structure allows for composition modifications across a wide chemical space, which in turn provides freedom
 in optimizing electronic and thermal transport properties. In this work, we explore one such variation: heterogeneous pnictogen substitution.
\begin{figure}[ht]
\scalebox{0.45}{\includegraphics{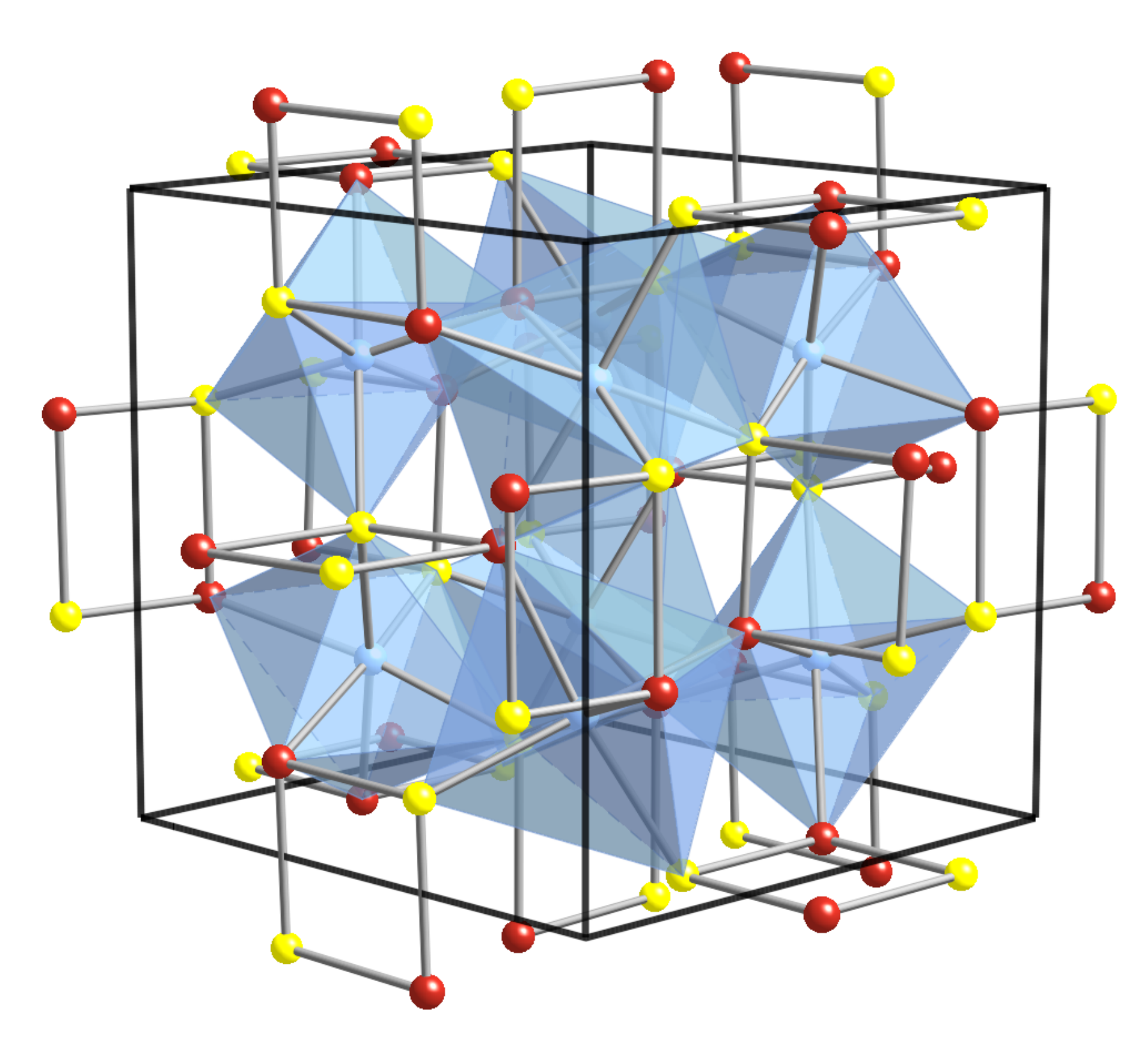}}
\caption{(Color online) Rhombohedral (R$\overline{3}$) unit cell of PSTS such as CoX$_{1.5}$Y$_{1.5}$ compound containing CoX$_3$Y$_3$ octahedra. Co centered octahedra linked by nearly rectangular X-Y 4-member rings (light gray lines), a characteristic feature of ternaries skutterudites. Black lines represent the unit cell (figure produced with CrystalMaker).}
\label{fig:skut}
\end{figure}

The conventional cubic unit cell of a binary skutterudite such as CoSb$_3$ (four formula units, space group n.
204) consists of a simple cubic transition metal  (M=Co) sublattice intertwined with square rings  formed by covalently bonded pnictogen (X$_4$=Sb$_4$) ions and
oriented  along  (100), (010) and (001) crystallographic directions. Each  transition
metal sits at the center of a distorted pnictogen  octahedron. In general, the six
pnictogens share nine electrons with the transition metals and two other
electrons with the two nearest pnictogen ions. Charge balance  constrains
the transition metal atom to have 9 electrons ($d^7s^2$) , thus  leaving a limited choice
of binaries with  Co, Rh or Ir as M  and P, As, Sb as X, in the absence of a filler ion.
Two large voids, that remain empty in the binary skutterudite cell, can be occupied by large
ions (fillers).\cite{sales_filled_1996,morelli_low_1995, nolas_effect_1996,shi_thermoelectric_2009,zebarjadi_effect_2010}
Substitutions and filling have a strong effect not only on the lattice thermal  conductivity but also
on the electronic band structure and associated transport  properties. This was pointed out both experimentally,\cite{fleurial_skutterudites:_1997} and by first principles band structure calculations.
\cite{singh_skutterudite_1994,feldman_lattice_1996,singh_calculated_1997,fornari_electronic_1999,feldman_lattice_2000,tritt_d._2001,lvvik_density-functional_2004,chaput_transport_2005,wee_effects_2010}

From a thermoelectricity point of view, binary skutterudites have comparatively large thermal conductivity, $k$. However, filling dramatically lowers the  lattice component of $k$, ($k_L$) and enhances ZT.\cite
{kjekshus_compounds_1974,lyons_preparation_1978,lutz_lattice_1981,slack_properties_1994,kim_structural_2010,caillat_novel_1994,borshchevsky_cosb3-irsb3_1994,caillat_properties_1996,borshchevsky_solid_1996,sales_filled_1996, morelli_low_1995,nolas_effect_1996, keppens_localized_1998}
Alloying on the transition metal site has also been explored as a strategy to
decrease thermal conductivity and control electronic transport.\cite{korestein__1977,rowe_g._1995,fleurial_thermoelectric_1995,caillat_properties_1996}
 However, the effect of chemical substitution on the
pnictogen site remains largely unexplored.

In this work we focus primarily on recently synthesized CoGe$_{1.5}$S$_{1.5}$, GeSn$_{1.5}$Te$_{1.5}$
and CoGe$_{1.5}$Te$_{1.5}$
where the substitution is  occurring on the pnictogen site of the
prototypical CoSb$_3$. We call  these materials pnictogen substituted ternary
skutterudites (PSTS). In order to obtain a complete comparison we also studied CoGe$_{1.5}$Se$_{1.5}$, CoSn$_{1.5}$S$_{1.5}$, and CoSn$_{1.5}$Se$_{1.5}$.
PSTS are experimentally observed to have a significantly lower thermal conductivity than CoSb$_3$,
\cite{vaqueiro_structure_2008,vaqueiro_structure_2006,nagamoto_thermoelectric_1997,vaqueiro_ternary_2008} and thus are attractive to
be investigated as potential thermoelectric materials. Until now the features of the band structure and of the phonon dispersion have
not been investigated in detail.

This paper is organized as follows. In Sec. II we briefly discuss the  first principles
methodology used to compute the electronic
and transport properties as well as the phonon dispersion. Sec. III is devoted to the main results on the  structural
features, Sec. IV contains the discussion on the electronic bands , Sec. V presents the phonon dispersions, and Sec. VI discusses the electronic transport coefficients. In Sec. VII we draw our conclusions.

 \begingroup
 \squeezetable
 \begin{table*}[t]
 \caption{Cystal structure of PSTS CoX$_{1.5}$Y$_{1.5}$ after relaxation of all internal degrees of freedom. The symmetry is R$\overline{3}$ (space group 148); experimental data (from Refs.\ \onlinecite{vaqueiro_structure_2008,vaqueiro_structure_2006,vaqueiro_ternary_2008,laufek_synthesis_2008,caillat_properties_1996,Schmidt_structure_1987})  are between parenthesis. CoSb$_3$ has higher symmetry (space group 204) but is treated as R$\overline{3}$ for easier comparison.}
\label{tab_pos1}
\begin{ruledtabular}
\begin{tabular}{lccccccc}
                &CoGe$_{1.5}$S$_{1.5}$ &CoGe$_{1.5}$Se$_{1.5}$&CoGe$_{1.5}$Te$_{1.5}$&CoSn$_{1.5}$S$_{1.5}$&CoSn$_{1.5}$Se$_{1.5}$ & CoSn$_{1.5}$Te$_{1.5}$& CoSb$_3$ \\ \hline
a$_L$ (\AA)     &  7.888 (8.010)    &    8.186         &     8.622 (8.699)  &  8.311  & 8.610  &  9.023 (9.122)&  8.972 (9.038) \\
$\alpha$ (deg) &  89.90 (89.94)    &    89.83          &     89.95  (89.99) &   89.87 &      89.98            &     89.97   (90.06)             &  90.0 \\
Co (2c) x         &  0.258 (0.258)    &  0.251     &   0.243 (0.247)   &  0.267 &   0.260  &   0.253 (0.250)   &    0.25     \\
Co (6f) x          & 0.258 (0.262)      &   0.253 &  0.249 (0.249)      &   0.262  &   0.260  &   0.255 (0.250)   &    0.25     \\ 
Co (6f) y          &  0.762 (0.755)    & 0.753      &  0.745 (0.745)   &  0.773  &   0.764  &   0.756 (0.750)   &    0.75     \\ 
Co (6f) z          &  0.754 (0.750)    & 0.752      &  0.747 (0.749)   &  0.758  &   0.755  &    0.751 (0.750)   &    0.75     \\ 
X$_A$ (6f) x    &  0.999 (0.000)    & 0.998     & 0.996 (0.995)    &   0.001 &   0.999 &   0.998 (0.998)     &   0.000   \\ 
X$_A$ (6f) y     &  0.335 (0.336)   &   0.327  & 0.318 (0.318)     &   0.333 &   0.328 &    0.321 (0.319)     &    0.334 (0.335)   \\ 
X$_A$ (6f) z     &  0.151 (0.148)   &  0.158  & 0.167 (0.166)      &   0.149 &   0.156 &   0.165 (0.162)    &     0.159 (0.158) \\ 
X$_B$  (6f) x     &  0.499 (0.498)  & 0.500    &   0.501 (0.501)    &   0.499 &   0.500 &  0.501 (0.500)     &    0.5    \\ 
X$_B$  (6f) y    &  0.835 (0.836)  & 0.827    &   0.818 (0.829)     &   0.834 &   0.828  & 0.821 (0.823)     &    0.834 (0.835)    \\ 
X$_B$  (6f) z    &  0.349 (0.350)  &  0.341    &   0.332 (0.338)    &   0.351 &   0.343 &0.335 (0.337)     &    0.341 (0.342)   \\ 
Y$_A$  (6f) x    & 0.000 (0.001)  & 0.00       & 0.999 (0.001)       &   0.001 &  0.001  &0.000 (0.001)     &    0.00      \\ 
Y$_A$  (6f) y    & 0.344 (0.347)  & 0.344     & 0.345 (0.346)      &    0.337 &  0.328  & 0.339 (0.338)     &   0.334 (0.335)     \\ 
Y$_A$  (6f) z    & 0.849 (0.856)   & 0.850    &0.851 (0.854)       &    0.840 &   0.843 & 0.845 (0.845)     &    0.841 (0.842)      \\ 
Y$_B$  (6f) x    &  0.502 (0.505)  &  0.503   &   0.505 (0.501)     &    0.501  &  0.502 &    0.503 (0.503)     &     0.5   \\ 
Y$_B$  (6f) y    &  0.844 (0.846)  &  0.844   &   0.845 (0.842)    &     0.837  &  0.838 &  0.839 (0.841)     &     0.834 (0.835)    \\ 
Y$_B$  (6f) z   &  0.650 (0.646)   &  0.649   &   0.648 (0.652)     &     0.659 &  0.657 &  0.655 (0.655)     &     0.659 (0.658)    \\

\end{tabular}
\end{ruledtabular}
\end{table*}
\endgroup

\section{Methodology}
All presented data are obtained by {\it ab initio} calculations within DFT formalism\cite{hohenberg_inhomogeneous_1964,kohn_self-consistent_1965} using the Perdew-Zunger LDA
exchange-correlation energy functional.~\cite{ceperley_ground_1980,perdew_self-interaction_1981} The effect of the core electrons is treated within the pseudopotential approach with both ultrasoft (Co, S),\cite{vanderbilt_soft_1990} and separable norm-conserving (Ge, Sn, Te)
pseudopotentials.  Plane-wave  basis was employed for the expansion of the valence electron wave functions and charge densities with the
kinetic-energy cutoffs of 30 Ry and 240 Ry, respectively.
All calculations were performed using a 4x4x4 Monkhorst-Pack k-point mesh to sample the Brillouin zone.
All internal atomic coordinates were relaxed within Broyden-Fletcher-Goldfarb-Shanno method until the forces on the nuclei were below $10^{-3}$ a.u. The theoretically optimized lattice provides a residual stress smaller than 5.8 KBar.

Phonons were computed using density functional perturbation theory (DFPT).\cite{baroni_phonons_2001} The dynamical matrix was Fourier interpolated on a fine {\bf q}-point mesh starting from a 2x2x2 grid.
All calculation were performed with the Quantum-ESPRESSO software.~\cite{giannozzi_quantum_2009}

Electronic transport coefficients are computed withing the BOLTZWAN code\cite{BOLTZWAN} using the Boltzmann transport equation (BTE) in the constant scattering
time approximation. Our methodology differs from other approaches (see for instance Ref. \onlinecite{madsen_boltztrap._2006}) in that we employ
 maximally localized Wannier functions (MLWF, Ref. \onlinecite{marzari_maximally_1997}) to map the first principles electronic structure on a tight-binding model
 and obtain band derivatives following follow the work of Yates et al. (Ref.\ \onlinecite{yates_spectral_2007}). The method is not sensitive to band
crossings and provides an efficient way to integrate Fermi velocities over the Brillouin zone.
The computation of MLWFs has been performed within the WANNIER90 package~\cite{mostofi_wannier90:_2008} using the Bloch states obtained
with the Quantum-ESPRESSO distribution. Relevant procedures for obtaining the band derivatives  are described in the Appendix and examples
of MLWF for CoSb$_3$ are shown in Fig. \ref{fig:CoSbwf}. We used also the BOLTZTRAP package\cite{madsen_boltztrap._2006} of Madsen and Singh to compare with 
previous calculations.

\section{Structural features}

 The two main structural units in prototypical CoSb$_3$ are transition metal centered pnictogen octahedra and pnictogen rings. In PSTS the symmetry decreases with respect to CoSb$_3$ and two different kinds of octahedra and rings can be identified. The structure of the pnictogen rings is known to have a strong influence on electronic bands, phonons, and consequently transport properties of binary and filled skutterudites.\cite{jung_importance_1990,llunell_electronic_1996,feldman_lattice_1996,wee_effects_2010}
The typical PSTS structure, MX$_{1.5}$Y$_{1.5}$  (space group n. 148)
is derived from the binary  counterpart by a substitution of  the  pnictogen ion with a
pair of elements from 14  (Ge, Sn) and 16 (S, Se, Te) groups.
The stoichiometry is preserved but  heterogeneity is introduced in the rectangular rings  in which the two different ions are opposite (trans) to each other.
The rhombohedral primitive cell contains 32 atoms and can be described as a corner sharing octahedral network
that contains two non-equivalent Co-sites (2c and 6f Wyckoff positions, 2c along the diagonal of the cube), two non-equivalent X-sites (6f and 6f), and two non-equivalent Y-sites (6f and 6f).
Each tilted octahedron is formed by group 14 and 16 ions ordered in alternating layer perpendicular to the [111] direction. In general, the pattern of Co off-center displacements is such that the structure is centrosymmetric.
 For all the three compounds we have analyzed, Co(2c) is off center in their respective octahedra and displaced  along the [111] direction toward the smaller surrounding ions; Co(6f) is also slightly displaced toward the smaller ions (this true for all cases except CoGe$_{1.5}$Se$_{1.5}$ where the covalent radii are very similar) but in more complex pattern compatible with the symmetry. The octahedral units are deformed and tilted (a$^+$a$^+$a$^+$ in Glazer notation). The tilting is established to form the bonds of the two non-equivalent four-member rings involving Y$_A$ and X$_A$ or with Y$_B$ and X$_B$ in the PSTS structure and involves a doubling of the unit cell with respect an ideal ReO$_3$ network.
Shorter bonds are formed along a preferred cartesian direction and, to accommodate the rigidity of the octahedra, longer bonds result in one of the perpendicular directions. The relative length of these bonds determines the deviations from the ideal
 square shape (Oftedal's law) of the pnictogen rings. In PSTS such deviations are larger than in CoSb$_3$ since the bonds have additional ionicity
 that tends to decrease the interatomic distances (Schomaker-Stevenson rule). The dihedral angle in the rings changes from $90.0 ^{\circ}$ in CoSb$_3$
to smaller values ranging from $81.7^{\circ}$ to $89.8^{\circ}$ degrees for all the compounds except CoSn$_{1.5}$S$_{1.5}$ and CoSn$_{1.5}$Se$_{1.5}$.
Our computed structural parameters are given in Tab.\  \ref{tab_pos1} and are within 2\% of the experimental data.
\cite{vaqueiro_structure_2008,vaqueiro_structure_2006,vaqueiro_ternary_2008,laufek_synthesis_2008,caillat_properties_1996,Schmidt_structure_1987}
The lattice
 parameter correlates well with the covalent radii of the main group elements and the cell remains pseudo-cubic with small rhombohedral angles.
Our data shows the expected correlation between the lattice parameter and the size of the substitution atoms on both pnictogen sites. For example,
among the Ge-substituted compounds CoGe$_{1.5}$S$_{1.5}$ has the smallest lattice size while CoGe$_{1.5}$Te$_{1.5}$ has the largest and the same trend also appears in
 the other substitution site.

\begin{figure*}[t]
 \subfigure[]{\includegraphics[scale=0.35]{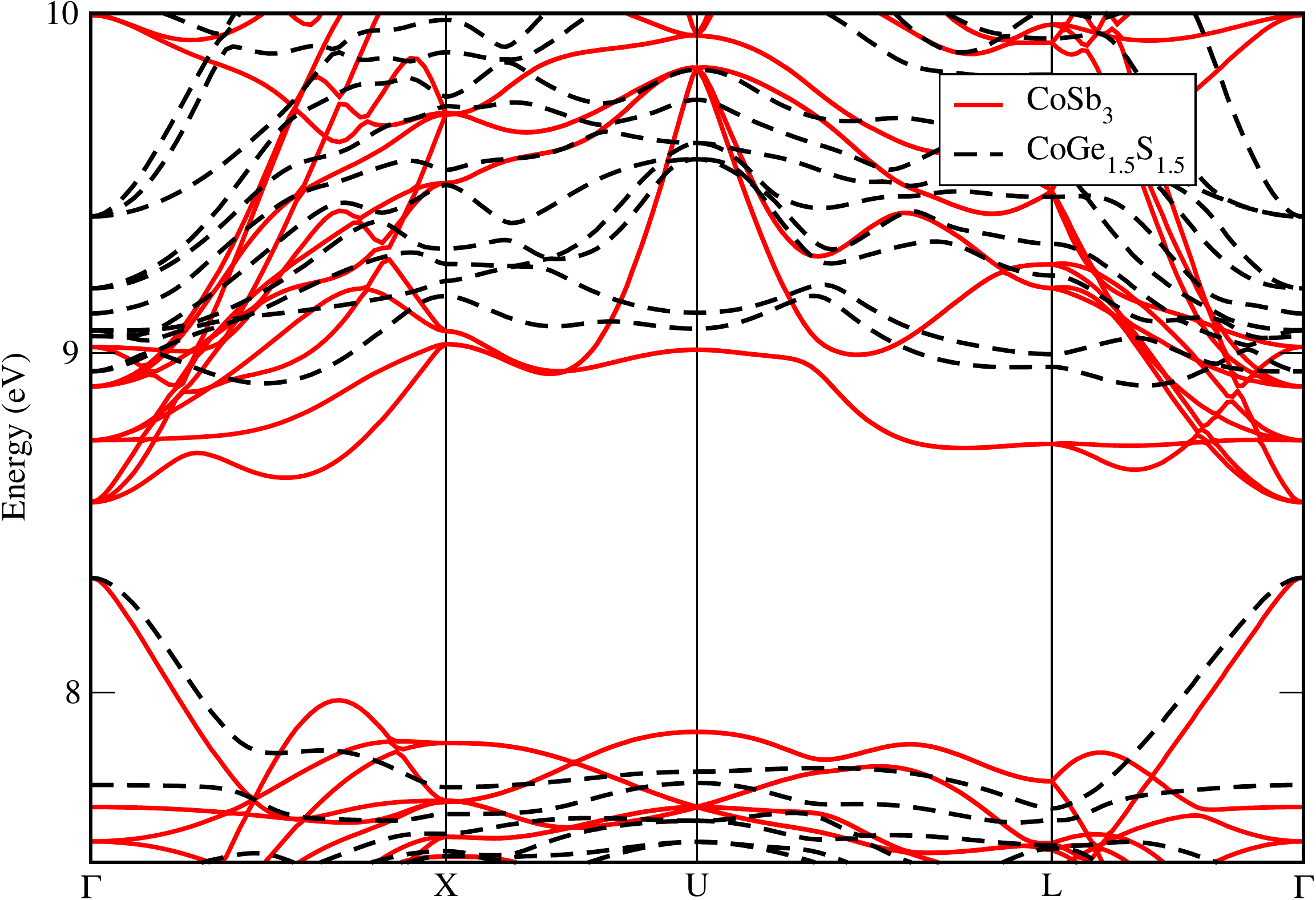}\label{fig:CoGeS_bs}}\hspace{0.5cm}
 \subfigure[]{\includegraphics[scale=0.35]{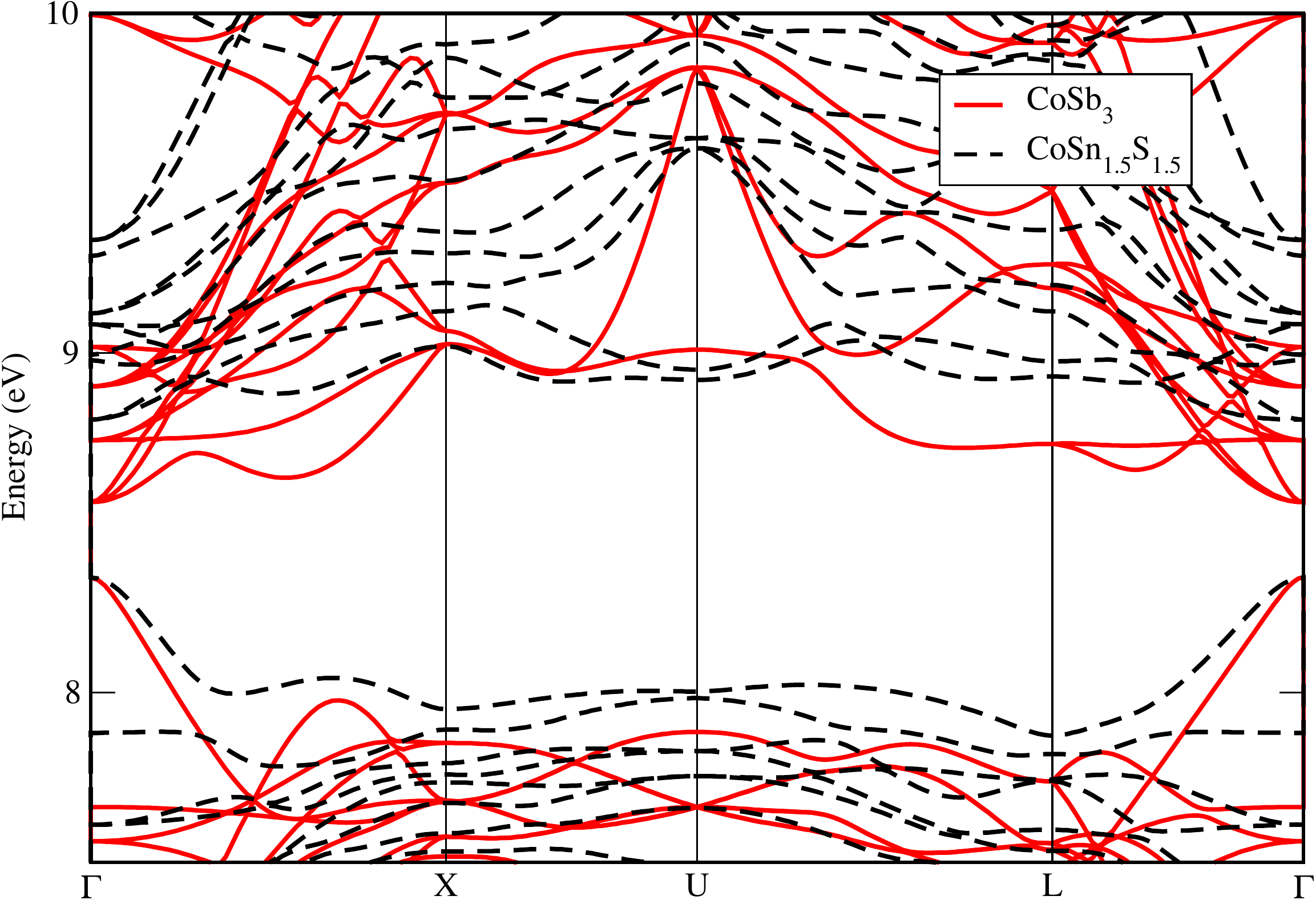}\label{fig:CoSnS_bs}}\\
\subfigure[]{\includegraphics[scale=0.35]{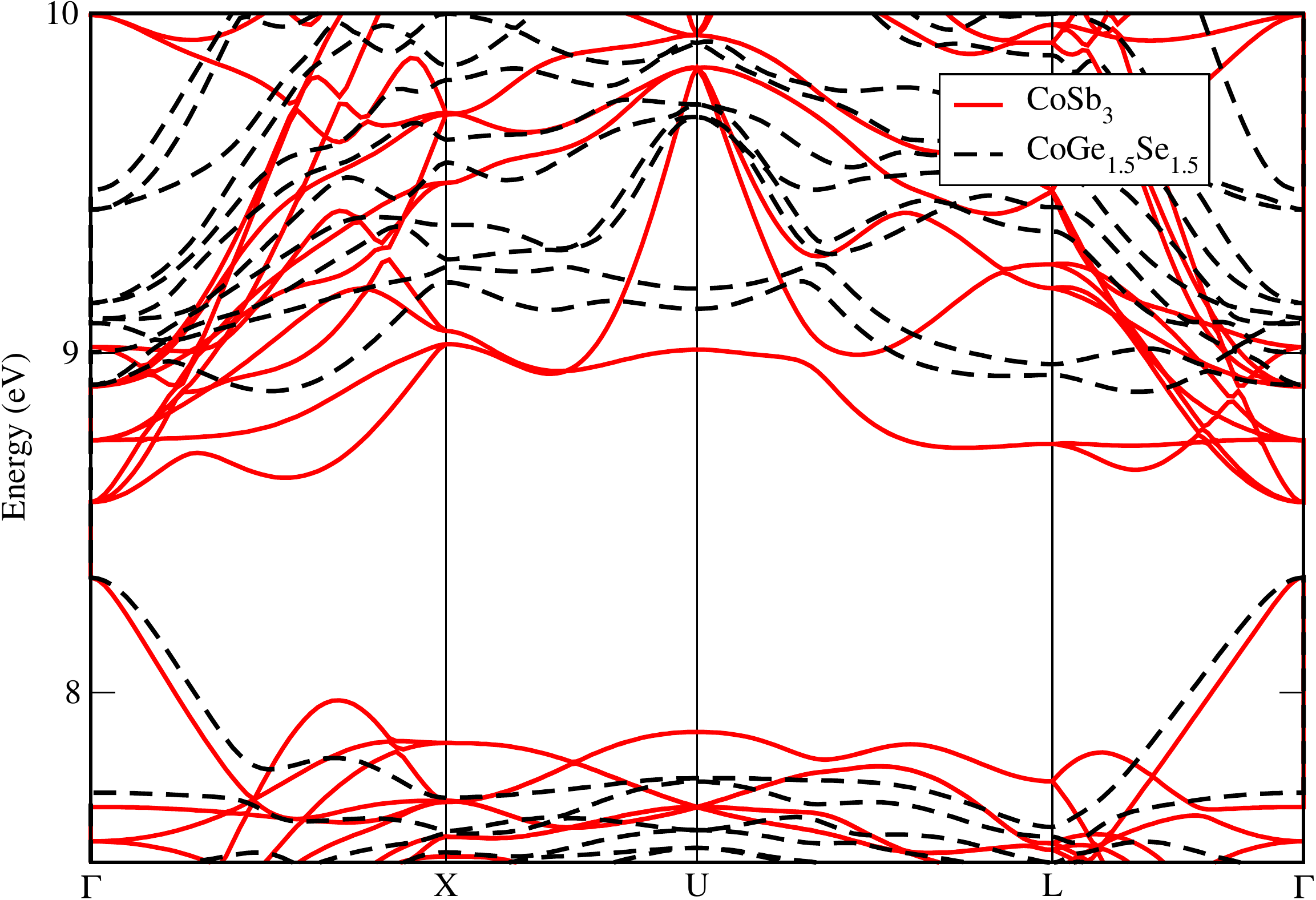}\label{fig:CoGeSe_bs}}\hspace{0.5cm}
 \subfigure[]{\includegraphics[scale=0.35]{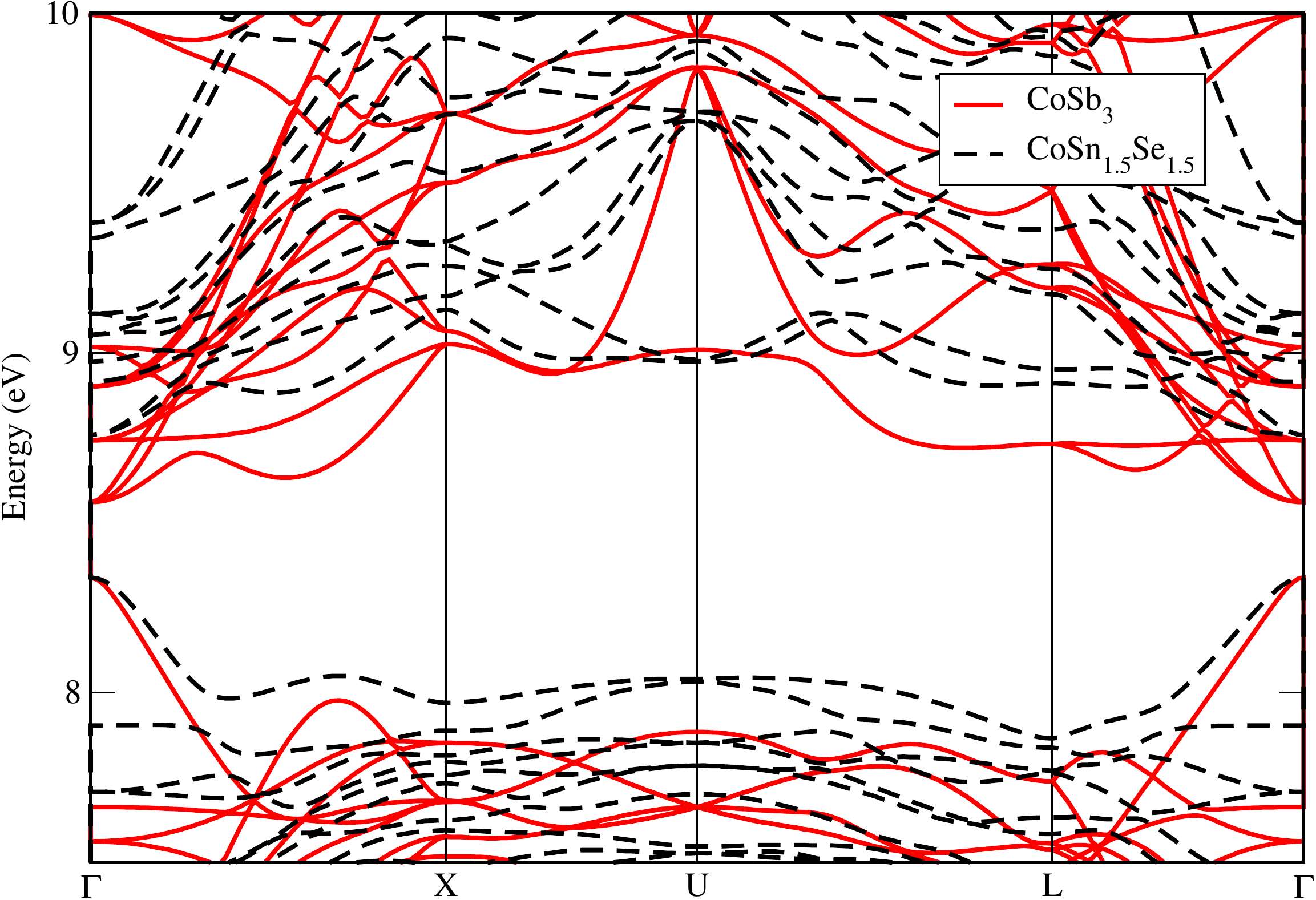}\label{fig:CoSnSe_bs}}\\
 \subfigure[]{\includegraphics[scale=0.35]{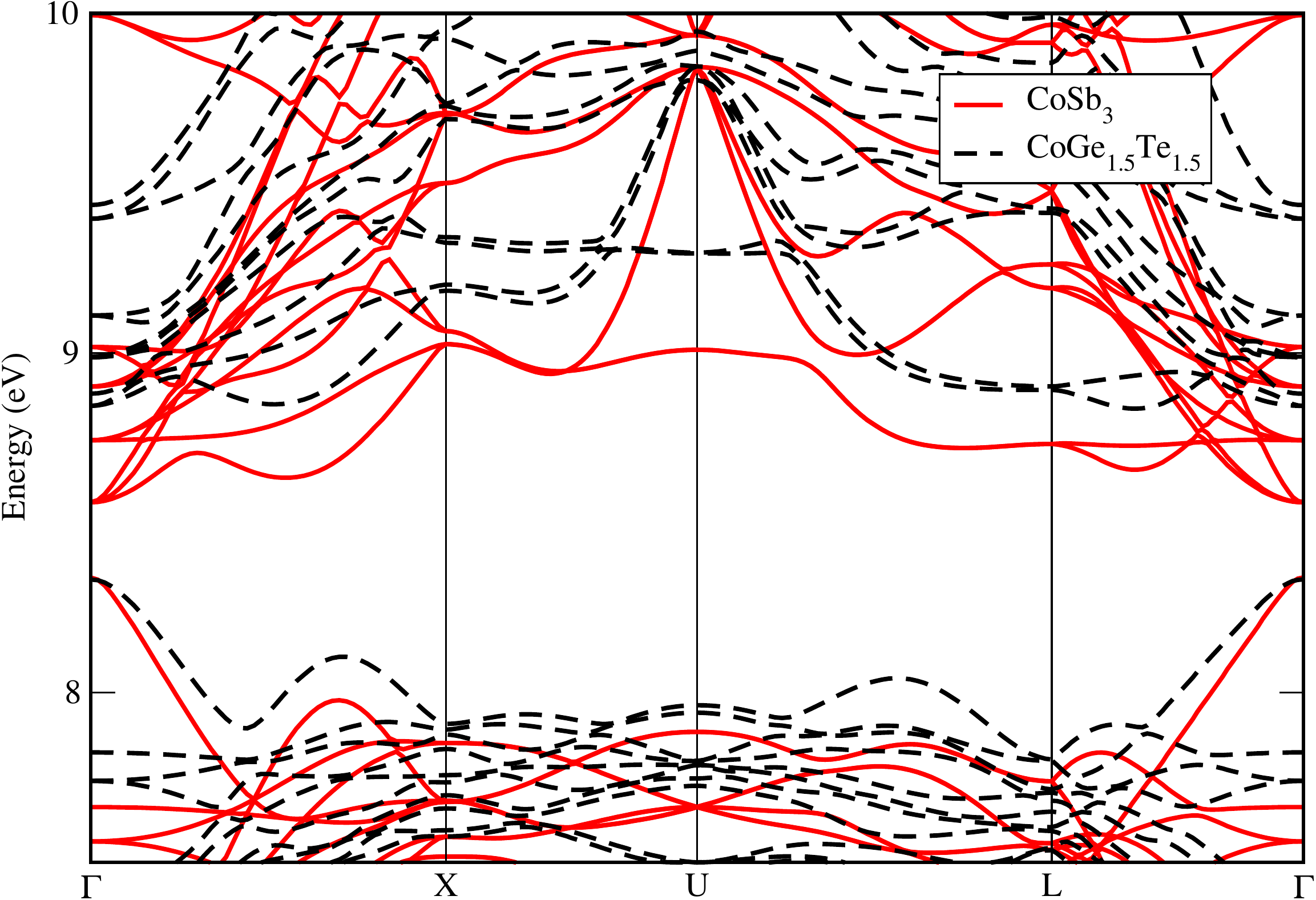}\label{fig:CoGeTe_bs}}\hspace{0.5cm}
 \subfigure[]{\includegraphics[scale=0.35]{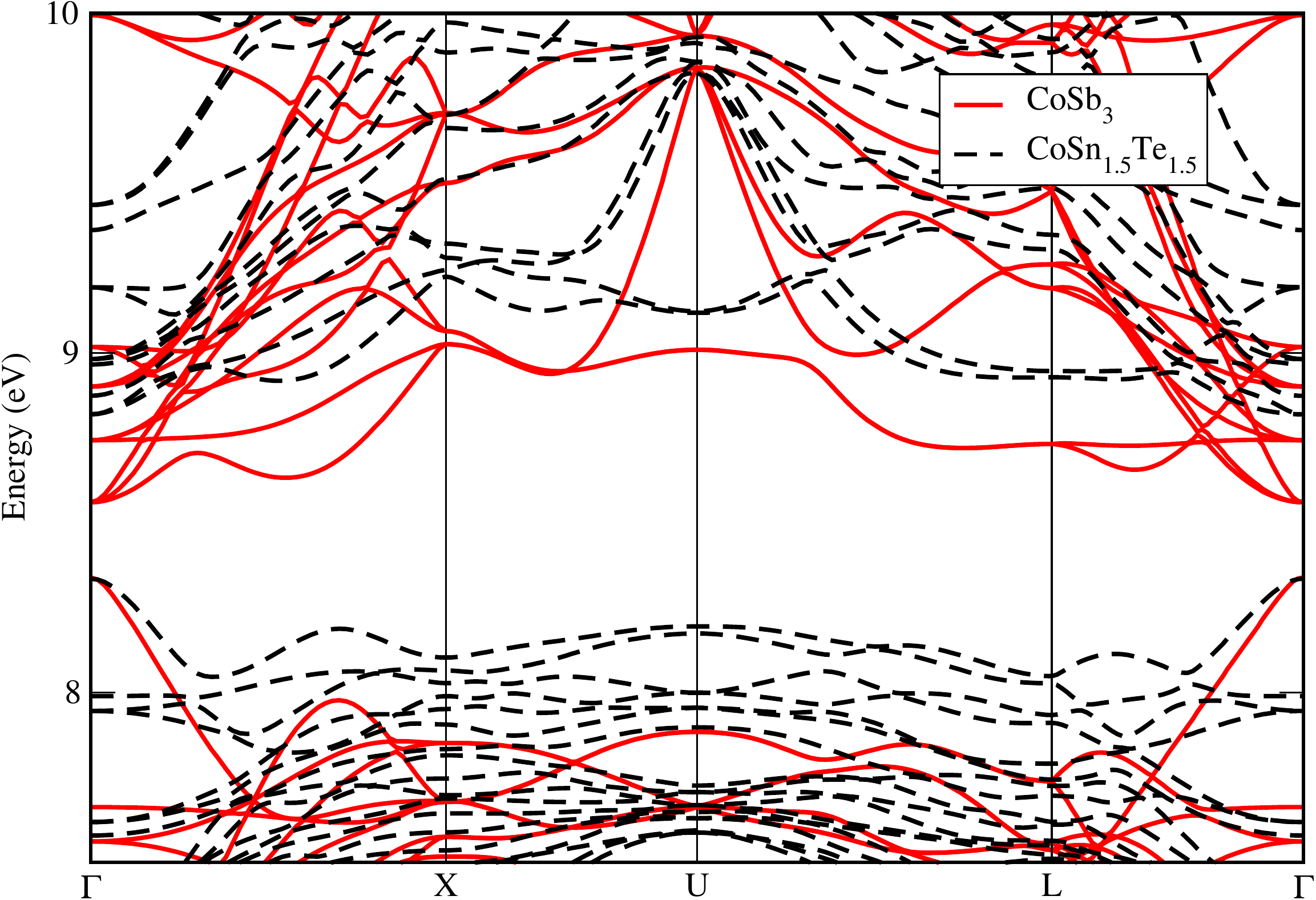}\label{fig:CoSnTe_bs}}\\
\caption{(Color online) First principles band structure of CoX$_{1.5}$Y$_{1.5}$ with (X,Y) = (Ge,S), (Ge,Se), (Ge,Te), (Sn,S), (Sn,Se), (Sn,Te) (dashed) compared to binary CoSb$_3$. CoSb$_3$ has been represented with R$\overline{3}$ symmetry for comparative purposes.}
\label{bs}
\end{figure*}

\section{Electronic Structure and Transport}
We calculate the electronic band structures (Fig. ~\ref{bs}) and compare them with the one of CoSb$_3$ (under equivalent symmetry representations) in order to investigate the effect of the pnictogen substitution. In all cases the valence bands consist of three separate manifolds. The lowest two are primarily derived from the unmixed $s$ states of two pnictogen types, and the splitting of the bands observed in PSTS is due to the different chemical nature and electronegativity of the pnictogen ions forming the rings. By comparison, in CoSb$_3$ the Sb-$s$ states contribute one single manifold. Both top valence and bottom conduction bands consist primarily of a mixture of Co $d$ states and pnictogen $p$ states with the majority of $d$ states lying below the top of the valence band.

Although the value of the computed band gap depends on the type of the exchange-correlation functional used,
~\cite{chaput_transport_2005,singh_skutterudite_1994,sofo_electronic_1998}
in all our cases the direct gap is 2-3 times larger than in CoSb$_3$  (it ranges from 0.41 eV in CoSn$_{1.5}$Se$_{1.5}$ to
0.61 in CoGe$_{1.5}$S$_{1.5}$). For comparison our calculations for CoSb$_3$ give
an energy gap of 0.22 eV in our DFT LDA calculations, while the experimentally measured values exhibit a wide variation.
~\cite{rakoto_shubnikovhaas_1998,rakoto_valence_1999,mandrus_electronic_1995,ghosez_first-principles_2007,sofo_electronic_1998,lefebvre-devos_bonding_2001,
kurmaev_electronic_2004,caillat_properties_1996,kawaharada_thermoelectric_2001}

Several effects contribute to the change in the band gap, mainly the  t$^*_{2g}$-e$^*_g$ derived manifold splitting and the
 flattening of the band dispersion in PSTS induced by the more ionic bonding.
Skutterudite systems typically possess a single band that disperses away from the t$^*_{2g}$ valence manifold and reaches
its maximum at $\Gamma$ point (the highest occupied band).
This band controls the lower edge of the energy gap and is important due to its role in transport in p-type materials
 because it provides carriers with small effective mass.
 In CoSn$_{1.5}$Te$_{1.5}$ the top of the valence band is about 170 meV above the low lying $d$-bands. This separation increases in
CoSn$_{1.5}$Te$_{1.5}$ (220 meV), in CoGe$_{1.5}$Se$_{1.5}$  (250 meV), and in the other PSTS, reaching values higher than in CoSb$_3$ (370 meV).
The second higher energy valence band (from the t$^*_{2g}$ manifold) of PSTS have a multivalley character with heavy effective masses. Particularly in CoGe$_{1.5}$Te$_{1.5}$ and in CoSn$_{1.5}$Te$_{1.5}$
the top of the valence band is relatively close in energy  to the bands below it; if this energy difference
could be further reduced the contribution from heavier carriers would enhance the p-type Seebeck coefficient favoring the thermoelectric performance. For comparison, in La filled CoSb$_3$
the first heavy valence band is about 70 meV below the top of the valence band due to an interaction between filler $f$-states and
the highest valence band.\cite{singh_calculated_1997}

In order to investigate the effects of ternary substitution on transport properties, we first evaluate the inverse of the hole effective
 mass tensor in the Wannier representation (see Appendix). The inverse of the effective mass is then defined as an average of the diagonal elements of the tensor, $1/m^*=\frac{1}{3}\sum_i{1/m_i}$. The corresponding values are 0.196$m_e$, 0.169$m_e$ and 0.134$m_e$ for CoGe$_{1.5}$S$_{1.5}$, CoGe$_{1.5}$Te$_{1.5}$ and CoSn$_{1.5}$Te$_{1.5}$ respectively, where $m_e$ is the electron mass.
These values are larger than reported $\approx$0.07$m_e$ for CoSb$_3$.
\cite{sofo_electronic_1998,caillat_properties_1996,arushanov_transport_1997}
In p-type PSTS samples higher effective masses of carriers are presumably responsible for the larger Seebeck coefficient values observed experimentally.
We find that the dispersion of the top valence band is also affected by the pnictogen substitutions; it is more parabolic than in CoSb$_3$
 although it also exhibits linear character of the dispersion close to $\Gamma$. This linearity has been suggested in earlier work to
affect hole transport and deviate from traditional semiconducting behavior.~\cite{singh_skutterudite_1994}

The lowest conduction energy levels also exhibit new features in PSTS. Several non-equivalent minima in $\Gamma-L$ and $\Gamma-X$ directions
 can provide pockets of carriers with large effective masses upon n-type doping.
This effect is also due to the decreased dispersion of pnictogen $p$ bands due to stronger ionicity.

\begin{figure}[b]
\scalebox{0.35}{\includegraphics{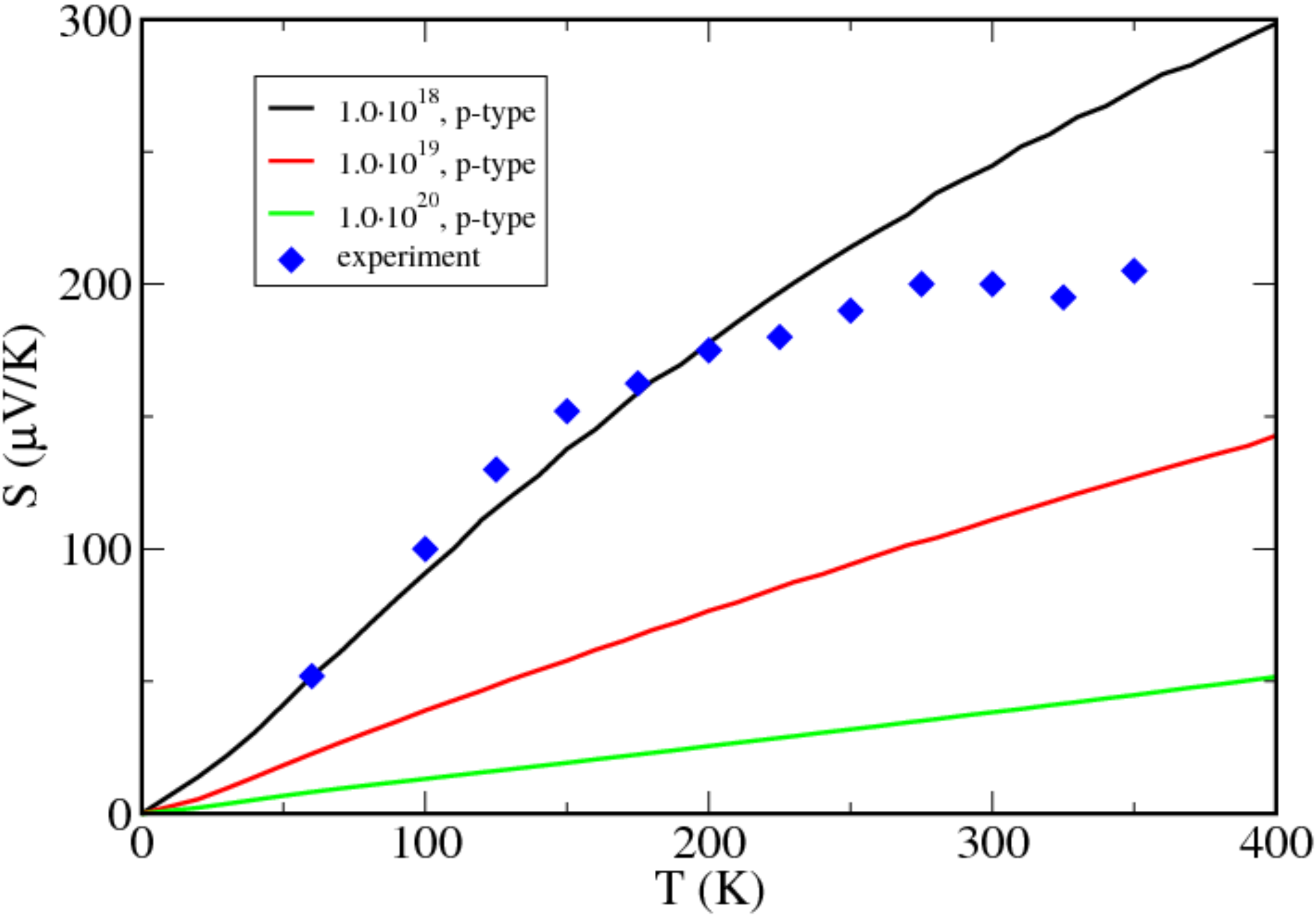}}
\caption{(Color online) Temperature and doping dependence of the Seebeck coefficient of CoGe$_{1.5}$S$_{1.5}$ compared to experimental intrinsic data.\cite{vaqueiro_structure_2008}  }
\label{fig:CoGeS_exp}
\end{figure}

\begin{figure}[t]
\scalebox{0.35}{\includegraphics{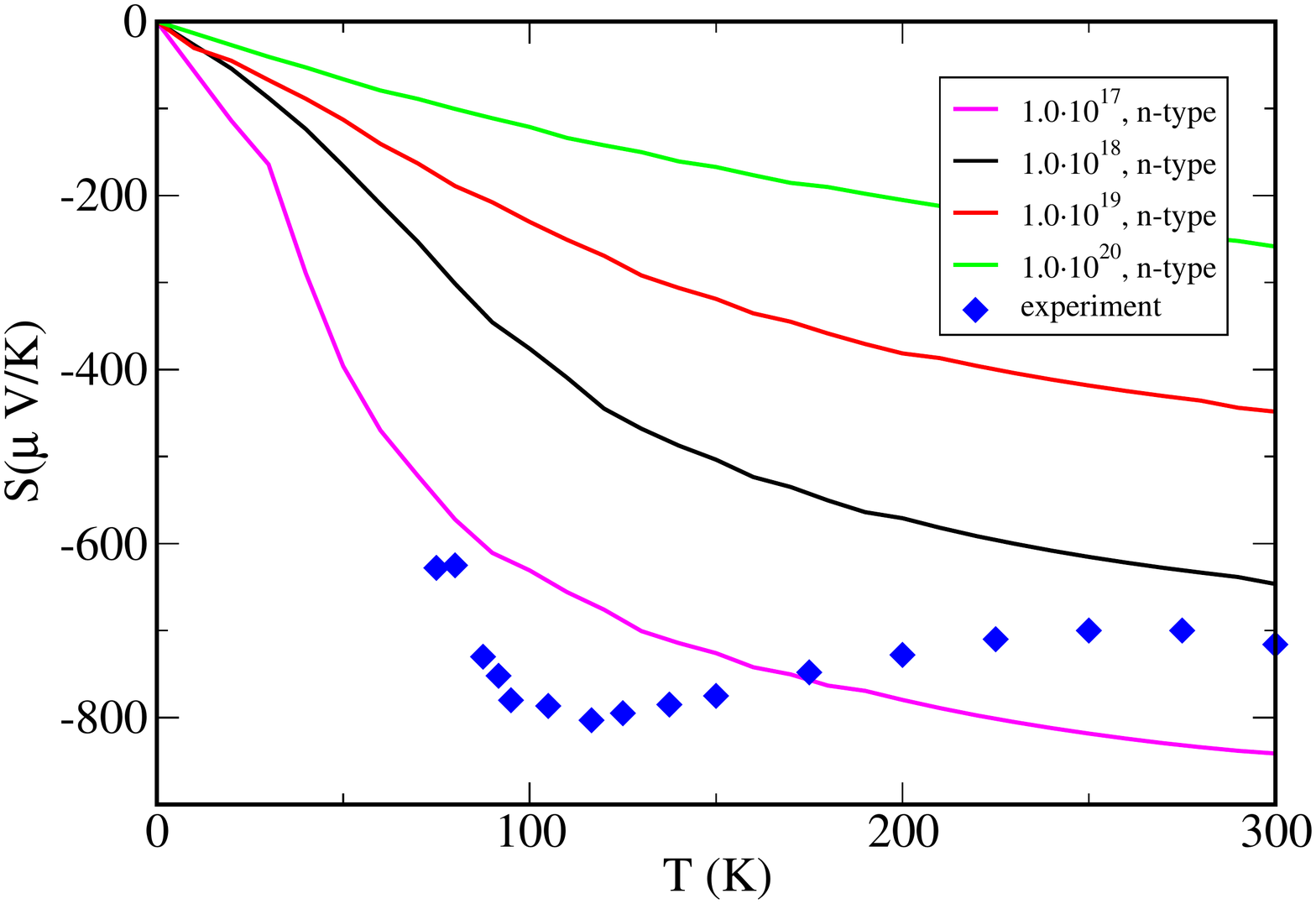}}
\caption{(Color online)  Temperature and doping dependence of the Seebeck coefficient of CoGe$_{1.5}$Te$_{1.5}$ compared to experimental intrinsic data~\cite{vaqueiro_structure_2006}.  }
\label{fig:CoGeTe_exp}
\end{figure}

\begin{figure}[t]
\scalebox{0.35}{\includegraphics{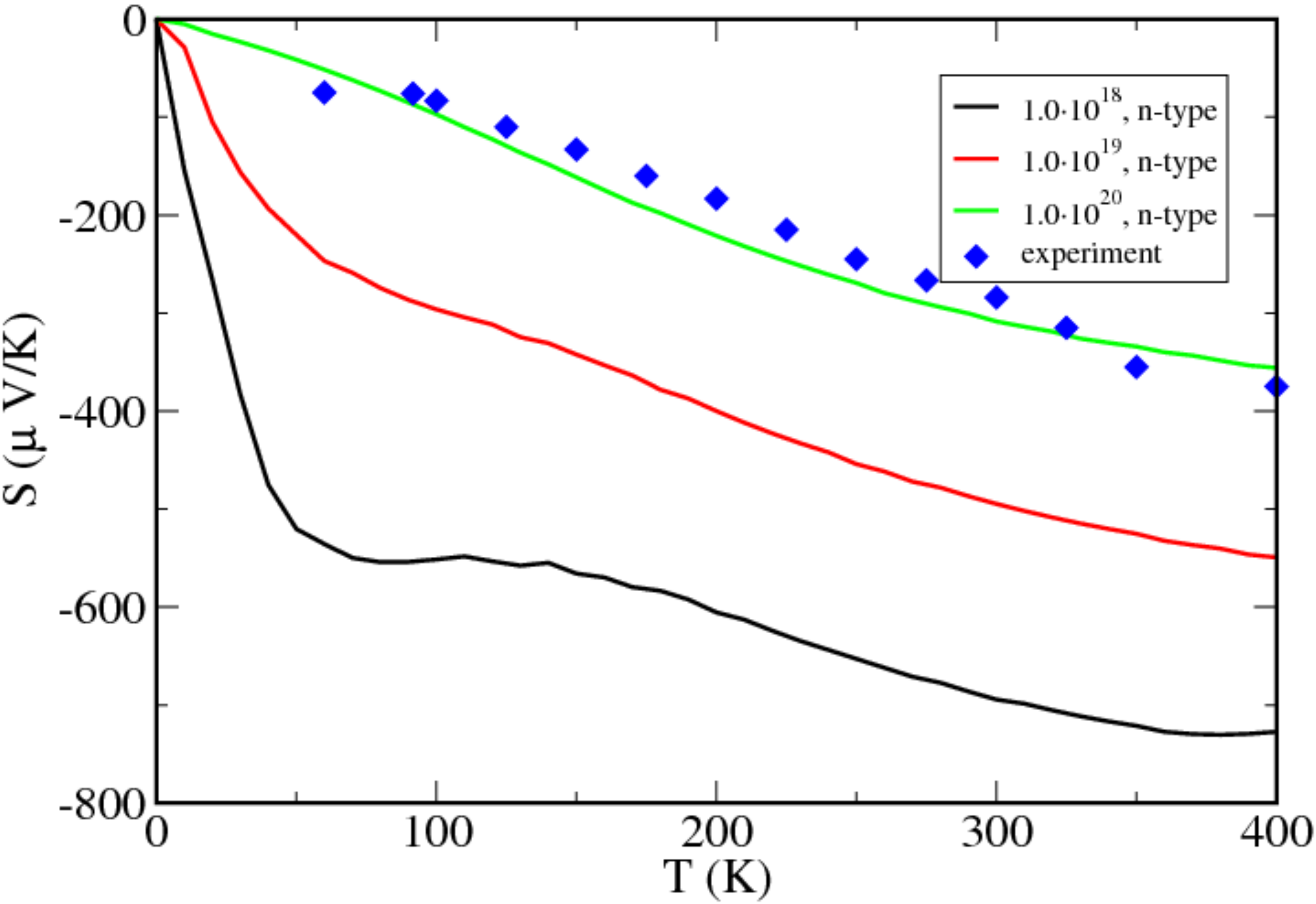}}
\caption{ (Color online) Thermopower of CoSn$_{1.5}$Te$_{1.5}$(bottom) compared to the experimental intrinsic data. \cite{vaqueiro_ternary_2008} Note the dip at low temperature as discussed in the text.}\label{fig:CoSnTe_exp}
\end{figure}

We derive the electrical conductivity and the Seebeck coefficient by solving the Boltzmann transport equation (BTE) in the constant relaxation
 time ($\tau$) approximation. We assume $\tau = 10$ fs in this work, which is commonly used for studying semiconductors.\cite{scheidemantel_transport_2003} This is an arbitrary choice since the scattering time for the PSTS is not known but it allows to establish the trends associated with band structure effects. Within the constant relaxation time approximation the Seebeck coefficient does not depend on $\tau$ and computed values can be compared directly with experimental data as we have done in Fig.~\ref{fig:CoGeS_exp}, \ref{fig:CoGeTe_exp}, and \ref{fig:CoSnTe_exp}.

\begin{figure}[ht]
\scalebox{0.33}{\includegraphics{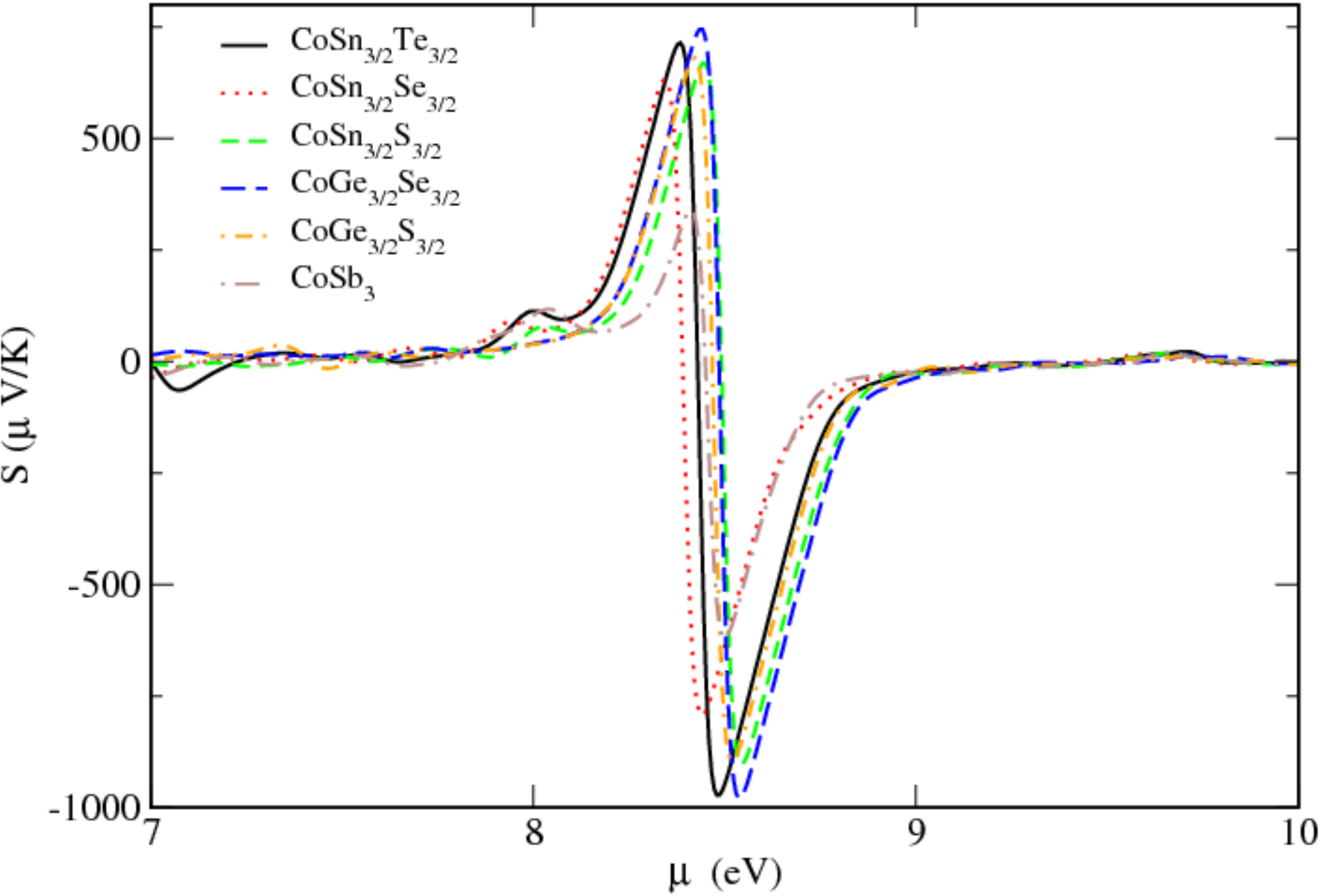}}
\caption{(Color online) Seebeck coefficients of ternary skutterudites at 300K as a function of the electron chemical potential (Fermi energy).} \label{fig:S}
\end{figure}

\begin{figure}[ht]
\scalebox{0.35}{\includegraphics{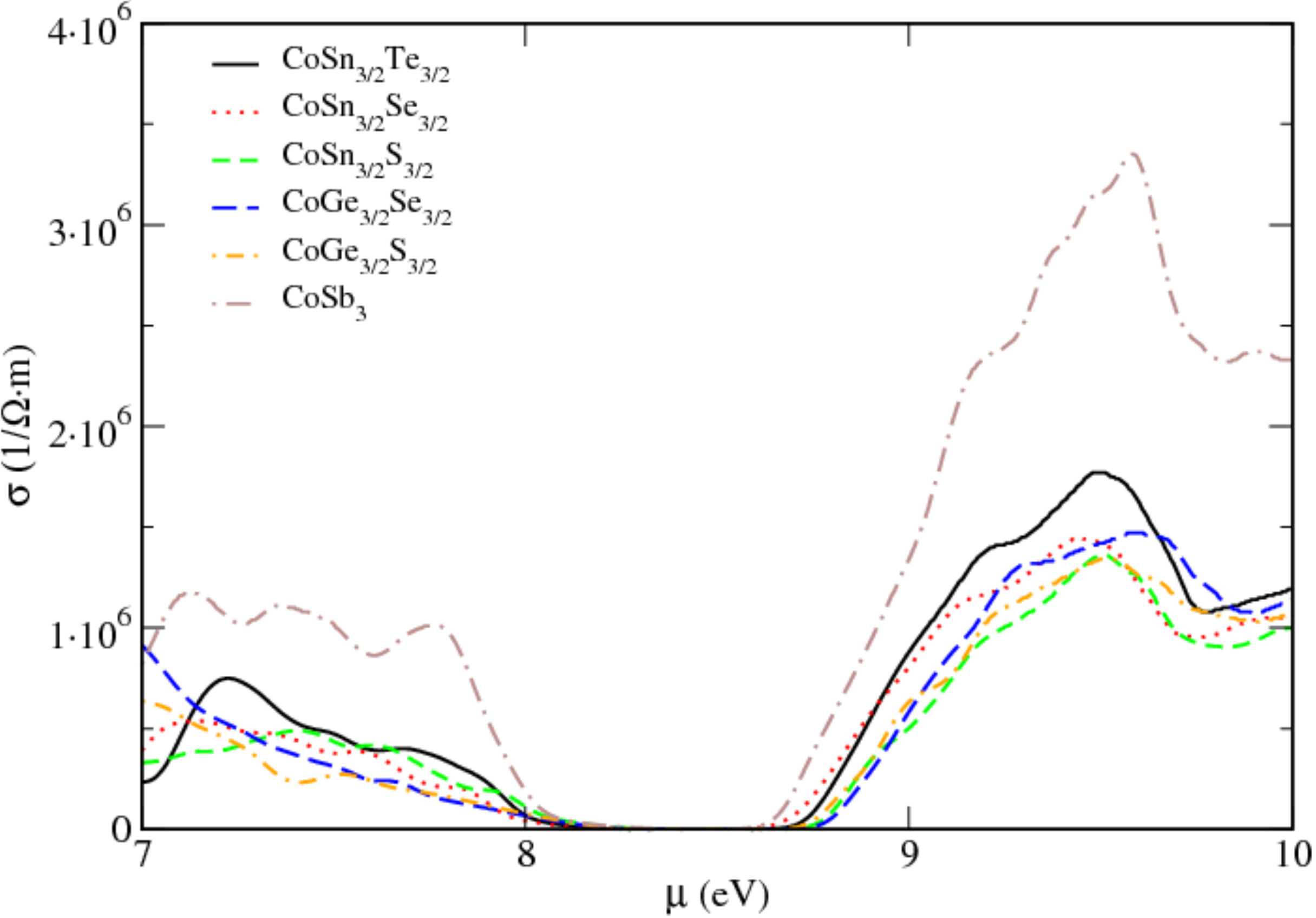}}
\caption{(Color online) Electrical conductivity of ternary skutterudites at 300K as a function of the electron chemical potential (Fermi energy).} \label{fig:cond_AnLi}
\end{figure}

\begin{figure}[ht]
\scalebox{0.33}{\includegraphics{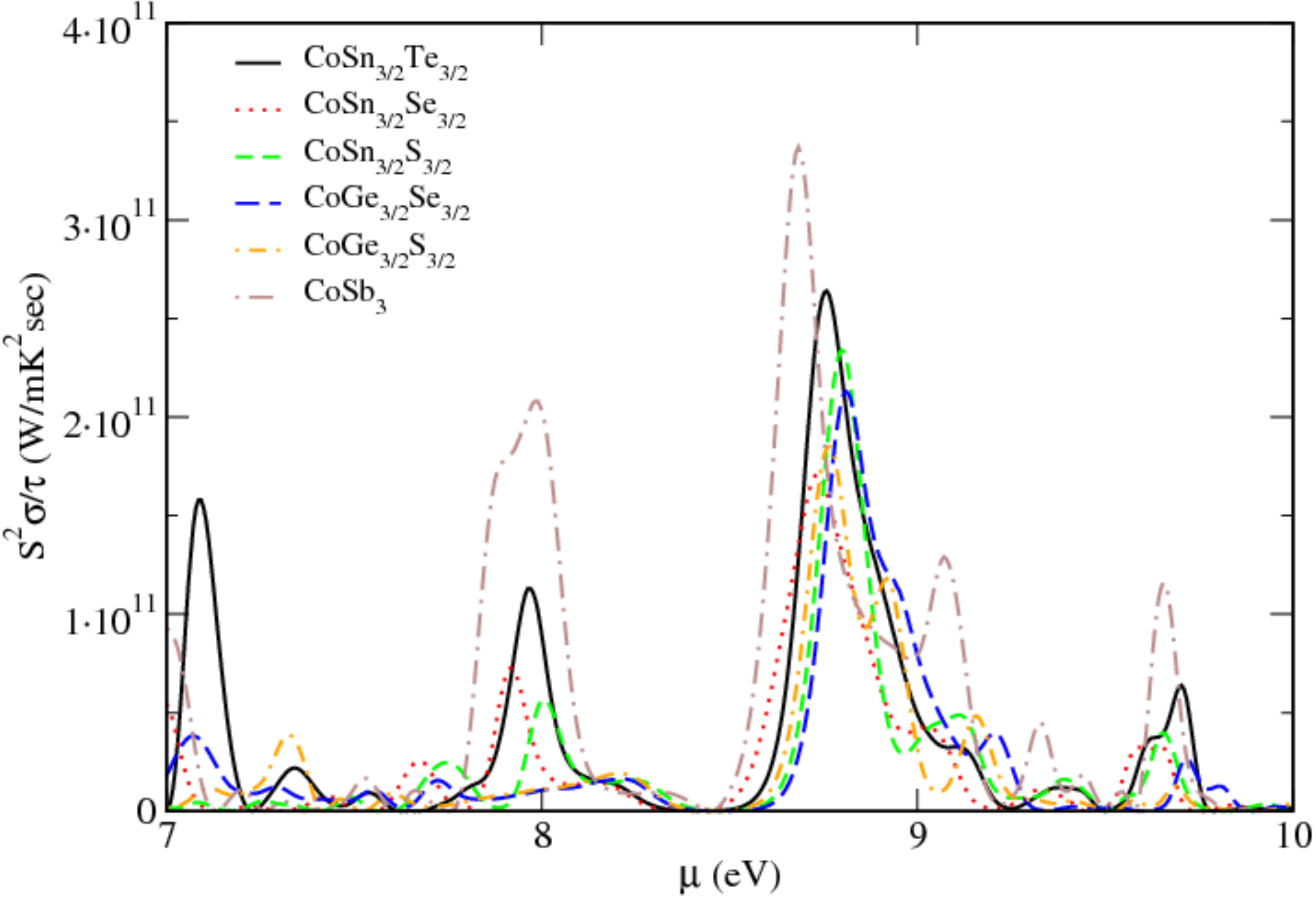}}
\caption{(Color online) Power factor of ternary skutterudites at 300K as a function of the electron chemical potential (Fermi energy).}\label{fig:power_AnLi}
\end{figure}

At the room temperature the Seebeck coefficient of CoGe$_{1.5}$S$_{1.5}$ ranges from 39 to 258 $\mu V/K$  for p-type doping in the range of 10$^{18}$ to 10$^{20}$ electron/cm$^{3}$.
Since the experimental value of the carrier concentration is not available for the samples under investigation we performed computations of the Seebeck coefficient in a range of possible carrier concentration by varying the position of the electron chemical potential. For carrier concentration of 10$^{20}$ holes per cm$^{3}$ our results agree with experimentally reported thermopower of CoGe$_{1.5}$S$_{1.5}$.
In the case of  CoGe$_{1.5}$Te$_{1.5}$ and CoSn$_{1.5}$Te$_{1.5}$, following the experimental findings, we tested n-type doping in the limits of 10$^{18}$ to 10$^{20}$ electron per cm$^3$.
At room temperature $S$ varies between -646 to -257 $\mu V/K$ for CoGe$_{1.5}$Te$_{1.5}$ and from -695 to -307 $\mu V/K$ for CoSn$_{1.5}$Te$_{1.5}$.
While computed values for CoSn$_{1.5}$Te$_{1.5}$ are in reasonable agreement with experimental data at a carrier concentration of 10$^{20}$ per cm$^3$, our results differ from the
experimental data for CoGe$_{1.5}$Te$_{1.5}$ at low temperatures. The Seebeck value reaches its minima of nearly -800  $\mu V/K$ at 115 K,
exhibiting an apparent dip and a subsequent increase decrease in the magnitude. It may be tempting to explain such a trend reversal by the bipolar effect,
i.e. decrease of $S$ due to thermal activation of minority carriers across the band gap. However, we argue that this feature derives only
from the electronic structure of the valence manifold at the experimental carrier concentration. The first reason is that the bipolar effect typically
sets in at temperatures where $k_B T$ is comparable to the band gap, and in PSTS the band gap is comparably large. The experimental values of $S$ for CoGe$_{1.5}$S$_{1.5}$,
 whose atomic and electronic structure is similar, exhibit the expected trend of increasing $S$ with temperature.
 The second and most compelling reason is the appearance of such a non-monotonic dip feature in the computed Seebeck coefficient temperature dependence of
 CoSn$_{1.5}$Te$_{1.5}$ (Fig. \ref{fig:CoSnTe_exp} ) at n=10$^{18}$ cm$^{-3}$. From the computed electronic band structure of CoSn$_{1.5}$Te$_{1.5}$
(Fig. \ref{bs}) we readily conclude that this non-monotonic feature is due to the intertwining and non-dispersive character of the valence band manifold.
 The deviation from the experimental behavior at higher temperatures is likely due to the the variation of the actual carrier concentration with
temperature and the inaccuracy of the constant $\tau$ approximation.\cite{caillat_properties_1996} At low temperatures impurity states
may also account for the observed strong dependence of transport properties on temperature.
\cite{dyck_effect_2002} For   CoGe$_{1.5}$Se$_{1.5}$,  CoSn$_{1.5}$Se$_{1.5}$,  and CoSn$_{1.5}$Te$_{1.5}$ we find values of $S$ to be between 200 and 400
 ${\mu V}/{K}$ for n-type doping and between 400 and 600  ${\mu V}/{K}$ at room temperature. The general trend, as shown in Fig. \ref{fig:S} is that the
 Seebeck coefficient in all six PSTS increases substantially with respect to CoSb$_3$ both for p-type and n-type doping; this
reflects the decreased band dispersion.

Experimental values of electronic resistivities at the room temperature are reported 30.6 $\Omega$cm, \cite{vaqueiro_structure_2008} for p-type CoGe$_{1.5}$S$_{1.5}$ ,
5.1 $\Omega$cm, \cite{vaqueiro_structure_2006}, for n-type  CoGe$_{1.5}$Te$_{1.5}$, and 0.33  $\Omega$cm, \cite{nagamoto_thermoelectric_1997} for
 n-type CoSn$_{1.5}$Te$_{1.5}$.
The theoretical results for conductivity at room temperature span over several orders of magnitude depending on the carrier concentration.
Our approach is to determine the carrier concentration as the one that produces the best match the temperature dependance of the thermopower to experimental measurements.

Selecting this doping level, we compute the room temperature electrical conductivity that is about two orders of magnitude larger than experimental values.
It must be noted that our methodology reproduces  the experimental results in wide range of temperature for the conductivity of CoSb$_3$  assuming $\tau = 2.5\ 10^{-14}$ (Ref. \onlinecite{wee_effects_2010}) when we use the experimentally determined carrier concentration values. Within the constant relaxation time approximation, where thermopower $S$ is independent of $\tau$, this approach provides a way to separate possible contributions to the discrepancy between theory and measurements. We can reasonably conclude that the features of the electronic structure alone are only partially responsible for much larger electrical conductivity with respect to experiment. Three quantities contribute to the electronic conductivity: effective masses, carrier concentration and scattering time $\tau$. The reasons for the discrepancy between experiment and theory could include inaccurate carrier concentrations and/or an anomalously short carrier lifetime. Impurity phases and defects in the experimental samples may also contribute to the discrepancy with the computed results.

In order to evaluate the potential of PSTS as active materials in thermoelectric devices, we compare their performance with the well-known CoSb$_3$ material. All PSTS have lower electronic conductivity than in a wide range of doping levels, as shown in Fig. \ref{fig:cond_AnLi}. Since the value of $\tau$ is taken to be the same, this reflects the larger band gap, decreased band dispersion and larger carrier effective masses. In Fig. \ref{fig:power_AnLi} we show the full power factors of all compositions as a function of doping level, and these results show a noticeably lower power factor for PSTS as compared with CoSb$_3$ in the p-type region and most of the n-type region. We conclude that, in the electronic transport aspect, PSTS are not likely to surpass the performance of CoSb$_3$-based systems, particularly for p-type materials, assuming the same carrier lifetimes. Furthermore, as we discuss below, carrier lifetimes in PSTS are likely reduced by the enhanced ionicity.


In an ideal crystal the scattering time includes contributions from the electron-phonon coupling, the larger contribution associated with deformation potential and Fr\"ohlich scattering.  The enhanced ionicity in PSTS suggest to consider effects associated to the Fr\"ohlich interaction. We have qualitatively analyzed
this contribution
 by evaluating the mode-resolved Born effective charges, defined by
\begin{eqnarray}
z^*_{\alpha}(\omega, {\bf k}) = {{\sum_{N,\beta}  Z^*_{N, \alpha \beta} e_{N, \beta}(\omega, {\bf k})}\over{\sqrt{\sum_{N,\beta} e_{N, \beta}(\omega, {\bf k})e_{N, \beta}(\omega, {\bf k})}
}}
\end{eqnarray}
to estimate for the polarization arising from the vibrational displacements and, consequently, the strength of electronic scattering.\cite{harrison_solid_1980}

Due to the smaller primitive cell and weak ionicity, CoSb$_3$ has a few (7) vibrational frequencies that exhibit non-zero $z_{\alpha}(\omega,{\bf k})$ at
the Brillouin zone center (see Sec. \ref{ph} for the phonon dispersions). In the low frequency region below 120 cm$^{-1}$ the mode resolved effective charges are less than 1 and the vibrational modes do not
effectively scatter electrons.
More significant scattering is expected when modes above 250 cm$^{-1}$,  with  $z_{\alpha}(\omega,\Gamma) \simeq 8$, become active.
 In PSTS the situation is quite different: we computed, in fact, many ``polar modes" with 2-3 time larger $z_{\alpha}$ than in CoSb$_3$.
These modes are distributed across the entire frequency spectrum.
This indicates that the enhanced polar scattering,  especially at low frequency, may affect strongly
 the electrical conductivity, as compared to CoSb$_3$. It is important to notice that the polar scattering contribution affects the thermal conductivity as well.

 \begin{table*}[ht]
 \caption{Transverse effective charges, Z*, computed with density functional perturbation theory. The full tensors are used to compute the electron-phonon polar scattering contribution but only $\frac{1}{3} Tr Z^*$ is reported here.}
\label{tab_zstar}
 \begin{ruledtabular}
\begin{tabular}{ccccc}
     &    $X=Ge$,$Y=S$         & $X=Ge$,$Y=Te$          & $X=Sn$,$Y=Te$     & $X=Y=Sb$  \\ \hline
Co (2c)    &        -5.140    &         -6.021    &     -5.914     &      -6.678   \\
Co (6f)    &        -4.895     &       -5.900     &     -5.810      &          --            \\
X$_A$       &        +3.227     &      +3.463     &     +3.506     &     +2.229              \\
X $_B$       &       +3.195     &      +3.462     &      +3.480     &        --              \\
Y$_A$        &     -0.035        &       +0.439      &    +0.334     &       --               \\
Y$_B$       &     -0.006        &       +0.430     &    +0.353    &         --             \\
\end{tabular}
\end{ruledtabular}
\end{table*}

\section{Phonons}\label{ph}

\begin{figure*}[ht]
\scalebox{0.70}{\includegraphics{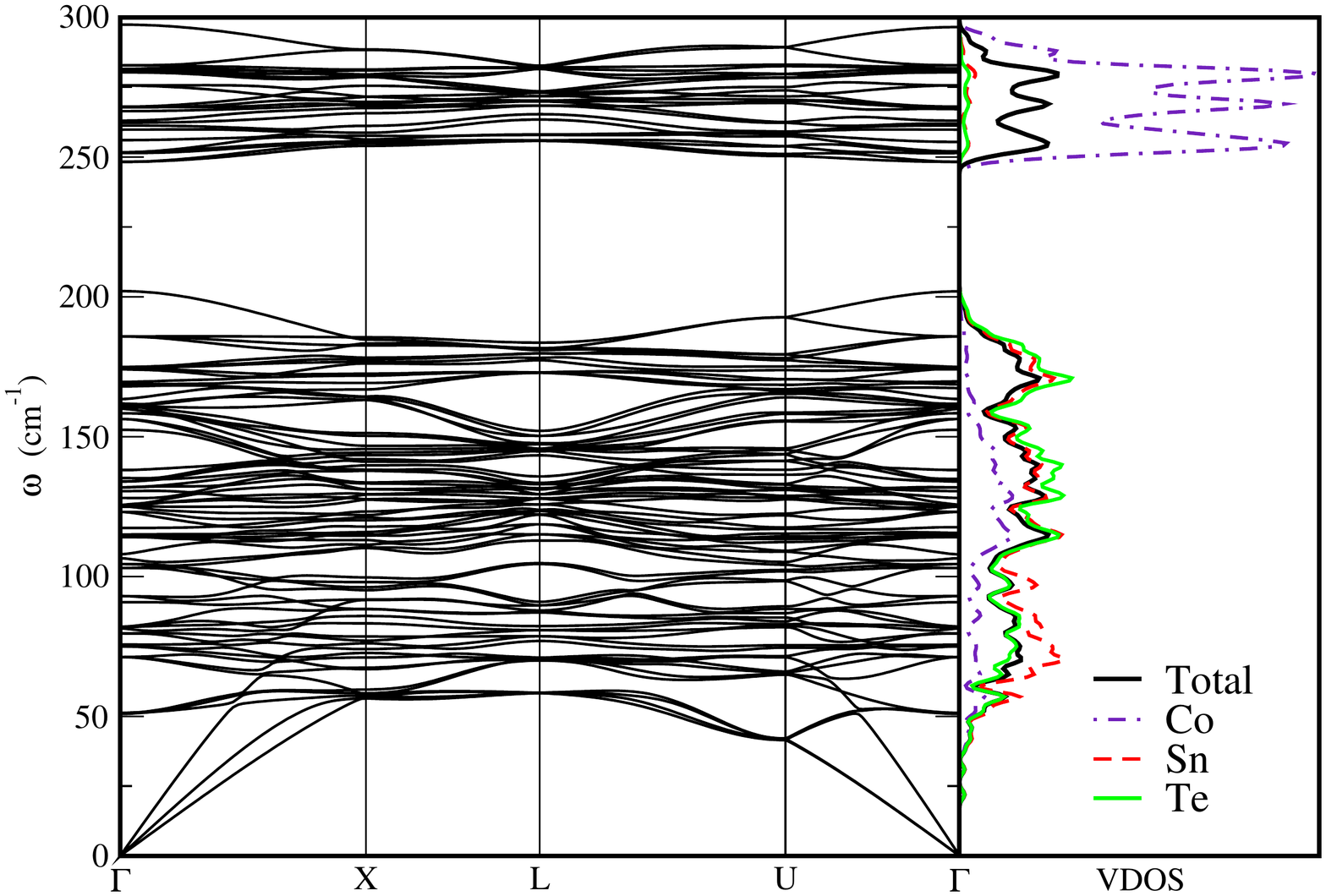}}
\caption{(Color online) Calculated phonon dispersion and atom projected vibrational densities of states of CoSn$_{1.5}$Te$_{1.5}$. }\label{fig:CoSnTe_ph}
 \end{figure*}

Full first principles phonon dispersion for filled and unfilled skutterudites were studied by Feldman et al.,\cite{feldman_lattice_1996,feldman_lattice_2000} Ghosez et al.,\cite{ghosez_first-principles_2007} and Wee et al.\cite{wee_effects_2010}
The vibrational spectrum of PSTS is an essential starting point to understand the role of the chemical substitutions in PSTS and develop models
for the low thermal conductivity observed in these materials. We present here the vibrational dispersions at the theoretically optimized structural parameters. For comparison, in CoSb$_3$ there are two main manifolds associated, respectively, with the vibration of the transition metal (between 250 and 300 cm$^{-1}$) and of the pnictogens (below about 200 cm$^{-1}$).
\begin{figure*}[ht]
\scalebox{0.70}{\includegraphics{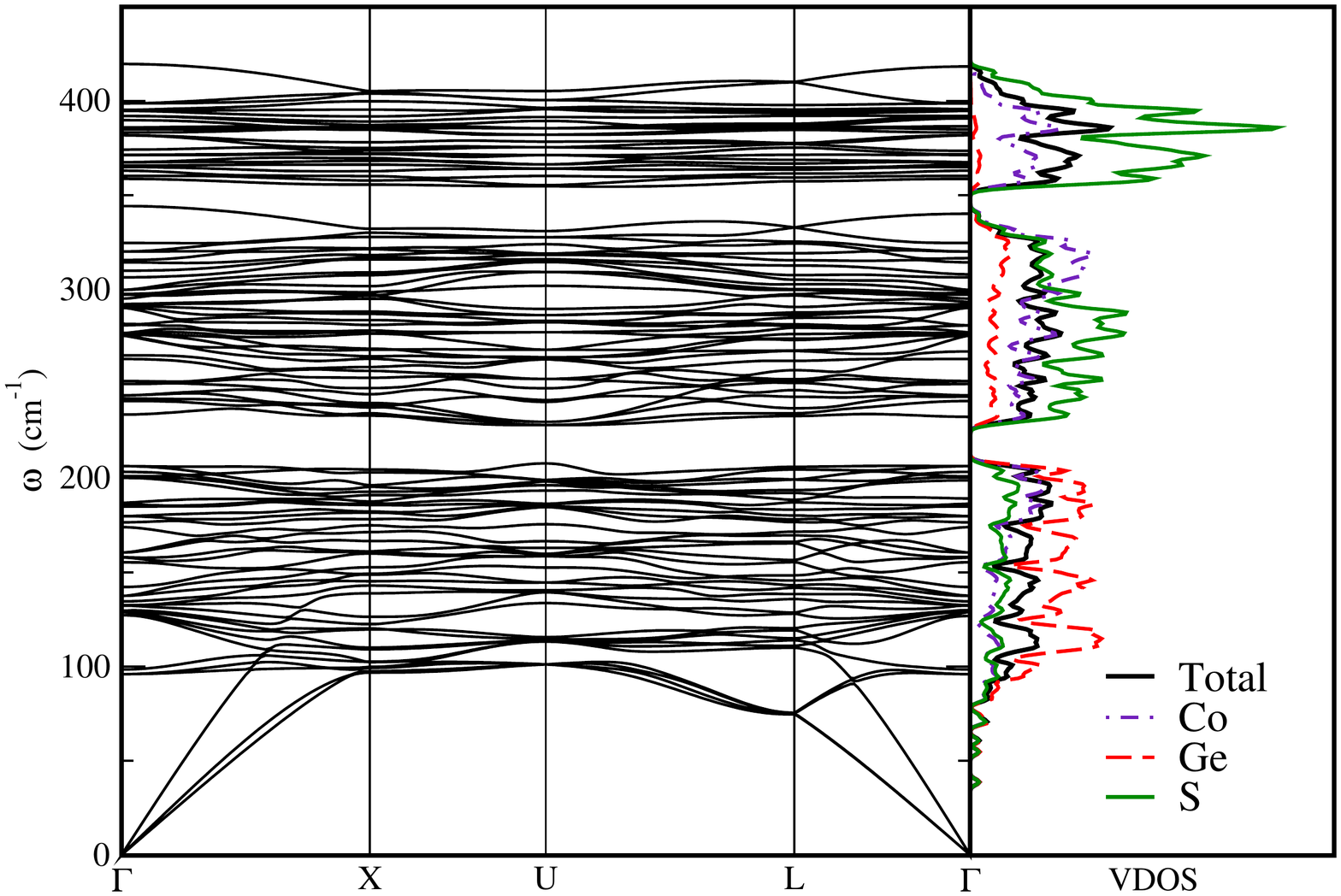}}
\caption{(Color online) Calculated phonon dispersion and atom projected vibrational densities of states of CoGe$_{1.5}$S$_{1.5}$.}\label{fig:CoGeS_ph}
\end{figure*}
In CoGe$_{1.5}$S$_{1.5}$   (Fig. \ref{fig:CoGeS_ph})
the comparable masses of Co and Ge result in the formation of vibrational modes that are mixed in character. The dispersion shows an additional manifold associated mainly with sulphur vibration above 350 cm$^{-1}$.  The motion of Co contributes across all frequencies with a larger contribution near 300 cm$^{-1}$. The frequency of the  lowest optical mode  (mainly Ge) at $\Gamma$ is at about 100 cm$^{-1}$ only slightly higher then the Sb-modes in CoSb$_3$. Similar features are observed in the dispersion of CoSn$_{1.5}$S$_{1.5}$  (not shown) where, of course, the Sn-derived modes extend to lower frequencies (about 75 cm$^{-1}$).
 \begin{figure*}[ht]
\scalebox{0.70}{\includegraphics{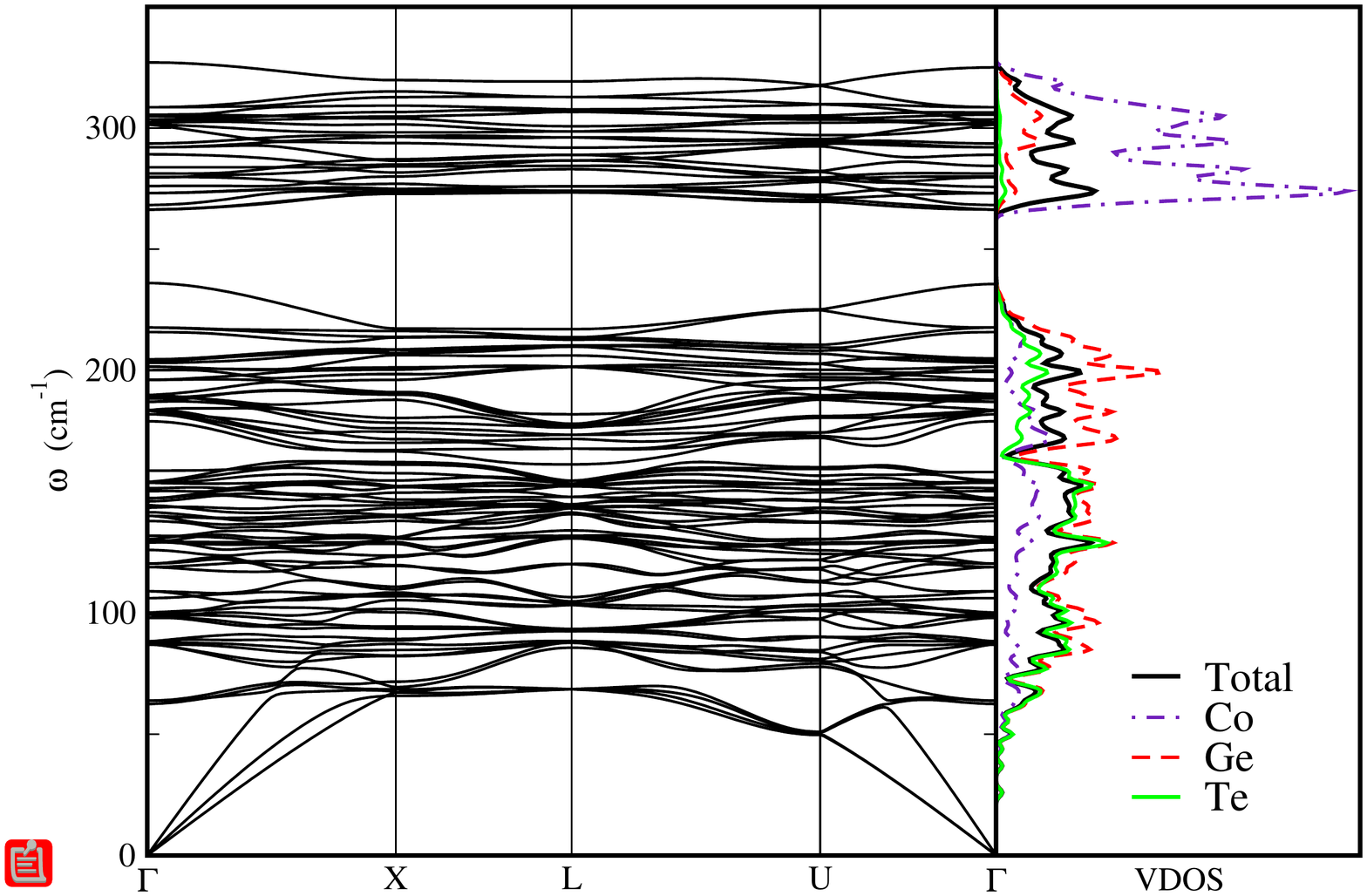}}
\caption{(Color online) Calculated phonon dispersion and atom projected vibrational densities of states of CoGe$_{1.5}$Te$_{1.5}$.}\label{fig:CoGeTe_ph}
 \end{figure*}
The  phonon dispersions of CoGe$_{1.5}$Te$_{1.5}$  (Fig. \ref{fig:CoGeTe_ph}) and CoSn$_{1.5}$Te$_{1.5}$ (Fig. \ref{fig:CoSnTe_ph}) exhibit two manifolds, similar to CoSb$_3$: the highest manifold is mostly from Co motion. The lowest frequency optical modes are at 65 cm$^{-1}$ in CoGe$_{1.5}$Te$_{1.5}$ and 50 cm$^{-1}$ in  CoSn$_{1.5}$Te$_{1.5}$. This is the frequency region where modes from filler atom vibrations are found in BaCo$_4$Sb$_{12}$ (Ref. \onlinecite{wee_effects_2010}) and may point to the phonon scattering channel responsible for the low thermal conductivity.

The group velocity of acoustic modes near $\Gamma$ determines the thermal conductivity and, in our calculations, correlates with the mass of the specific pnictogen substituted ions.
It is interesting to notice that the sound velocities in CoSn$_{1.5}$Te$_{1.5}$ are very similar to those of CoSb$_3$. In other PSTS we found values higher than those of CoSb$_3$. Based on these results and the overall phonon dispersions, it is reasonable to argue that scattering phenomena differ substantially between CoSb$_3$ and PSTS probably due to the different character of the bonding in the rings. Phonon dispersions alone cannot explain the low thermal conductivity values of observed experimentally for PSTS. More work in the direction of understanding the anharmonic scattering in these materials is required.

\section{Conclusions}

We discussed structural aspects, electronic structure and transport, and phonon dispersions of pnictogen substituted ternary skutterudites (PSTS).
These materials are potentially interesting for thermoelectric applications due to the exhibited low lattice thermal conductivity. Unfortunately the electronic
transport is not as favorable because of the low electrical conductivities.

We justified the large Seebeck coefficients by analyzing the electronic bands structures: a decreased dispersion compared with CoSb$_3$ as well as a  multivalley character with heavy carrier effective masses. The values of electronic conductivity are lower than for CoSb$_3$  and have a strong dependence upon carrier concentration. We explored the upper limits on the power factor of PSTS in a wide range of carrier concentrations
and found that they are unlikely to surpass those of CoSb$_3$. More effort should be invested in understanding the reasons for low measured values and find way to increase electronic conductivity in these materials.

\section{Acknowledgments}

The authors are grateful to Prof. Z. F. Ren and Prof. G. Chen for valuable discussions. This work was carried out as part of the MIT Energy Initiative, with financial support from Robert Bosch LLC. Additional funding was provided by the NSF-DOE Partnership in Thermoelectrics (CBET-0853350).

\begin{figure}[t]
  \label{fig:CoSbwf}
  \subfigure[]{\includegraphics[scale=0.25]{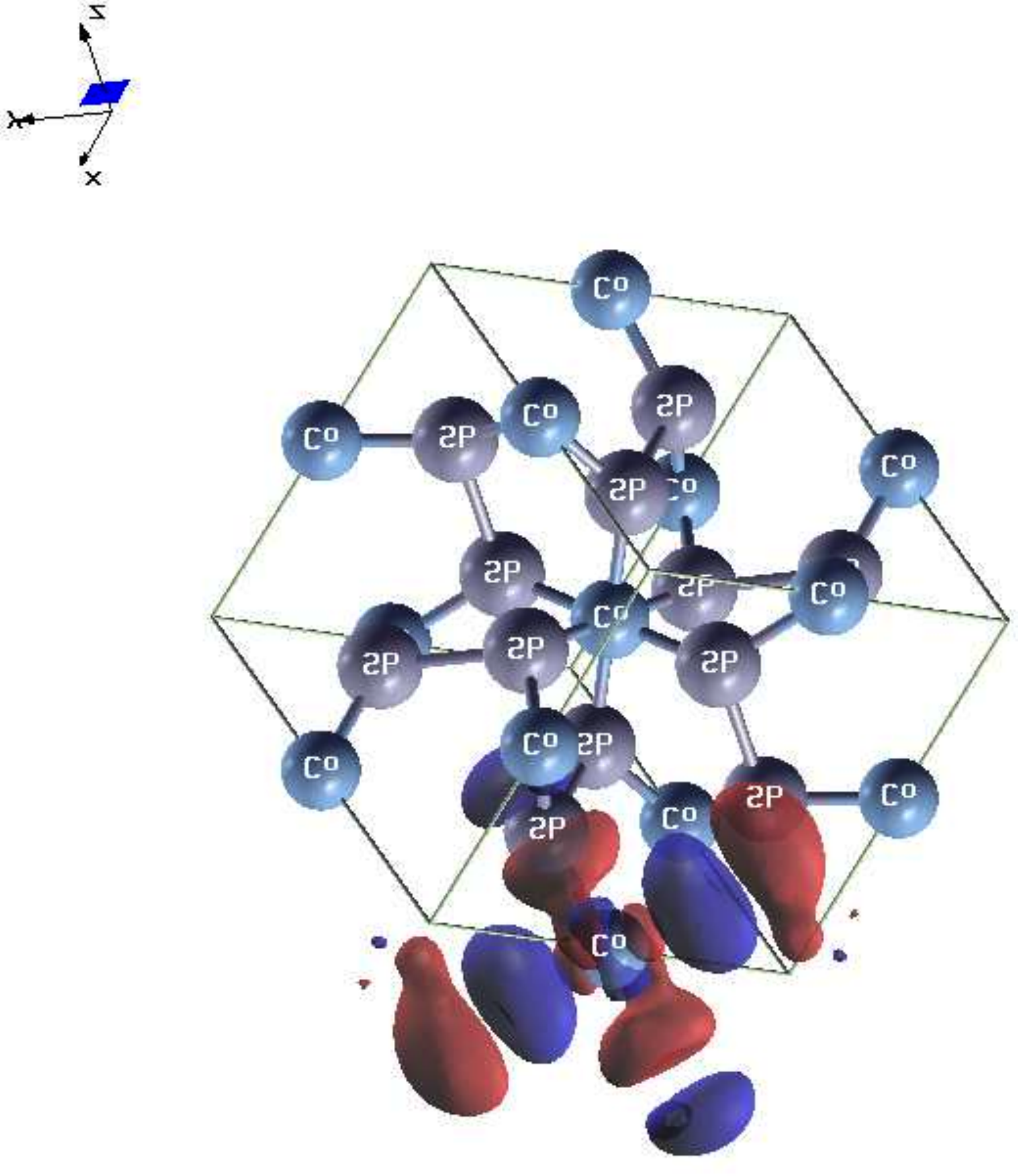}}\quad
  \subfigure[]{\includegraphics[scale=0.19]{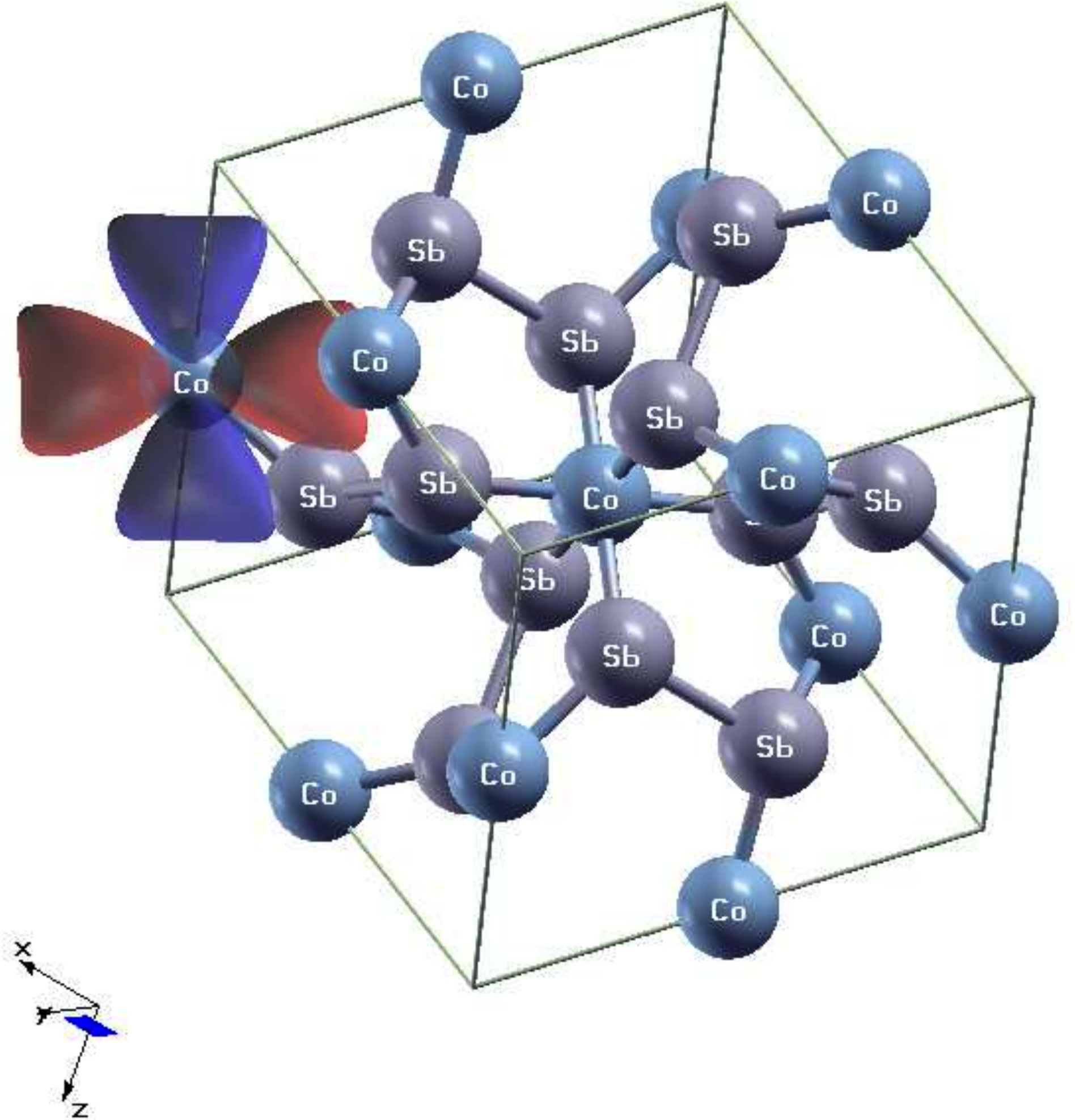}}\\
  \subfigure[]{\includegraphics[scale=0.18]{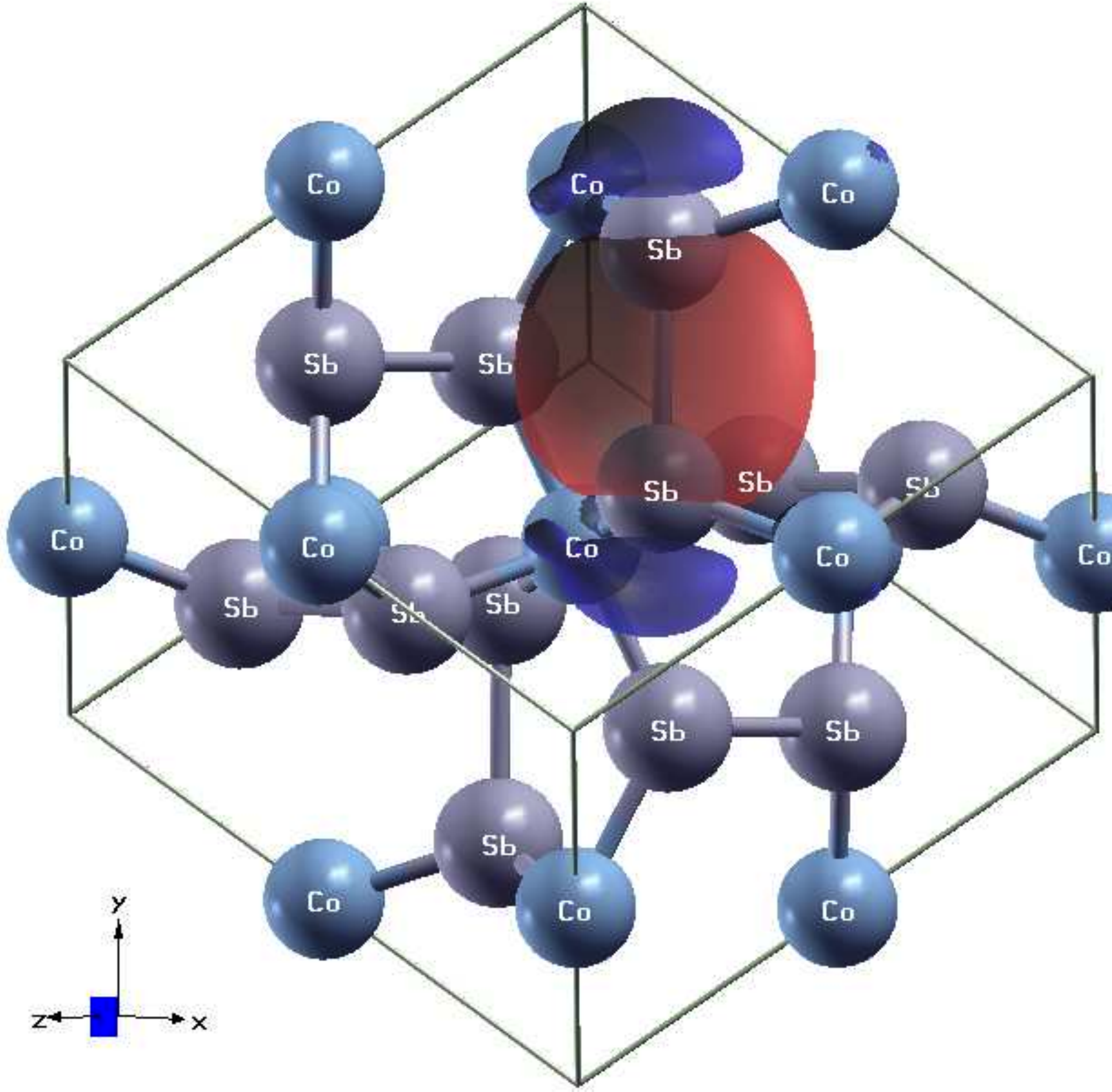}}\quad
 \subfigure[]{\includegraphics[scale=0.18]{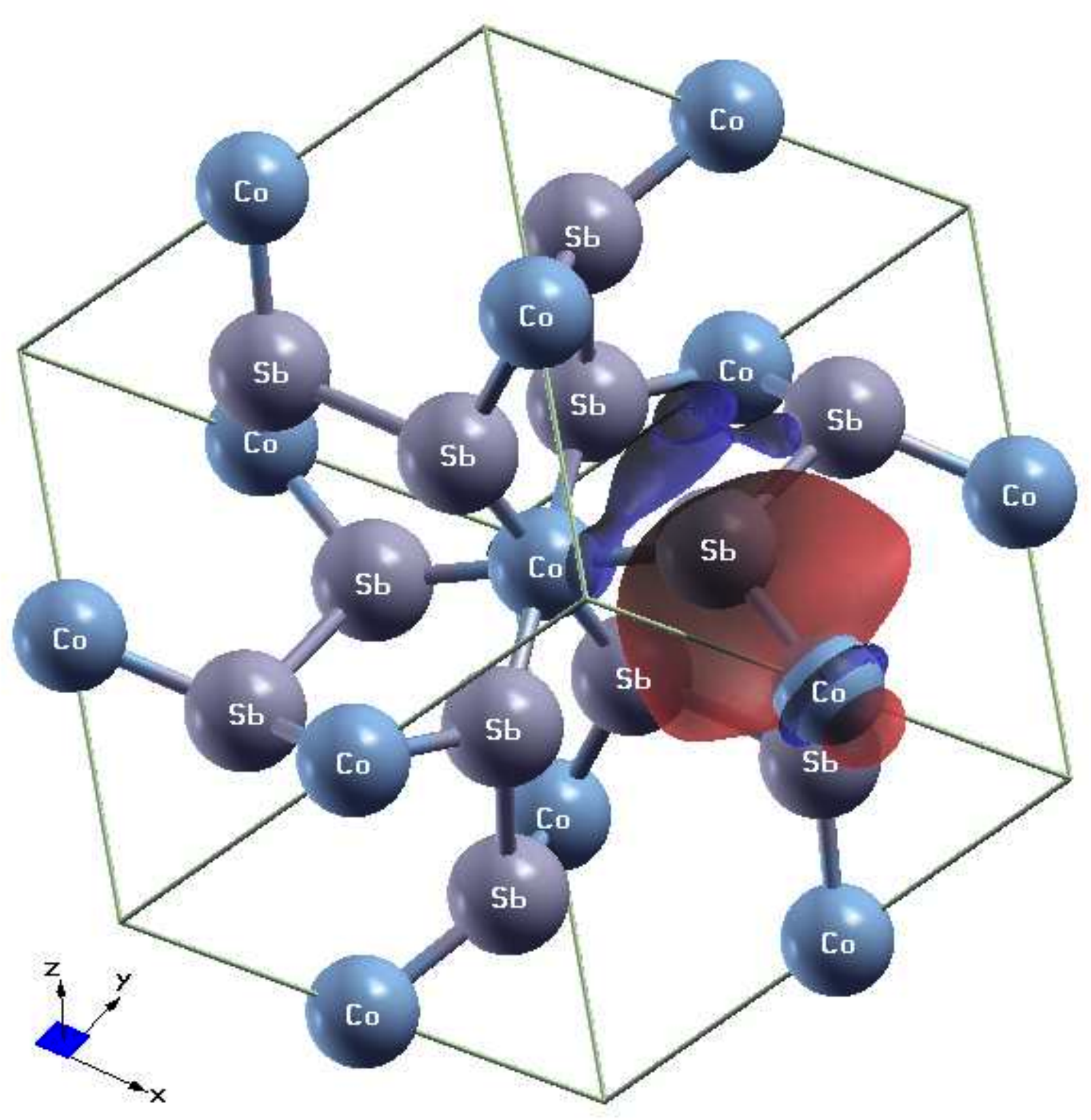}}\quad
\caption{(Color online) Contour-surface plots of the Maximally Localized Wannier Functions for: a) conduction antibonding states, b)  occupied Co $t_{2g}$ states, c) Sb-Sb bonding states,
d) $Sb-Co$ bonding states. Oppositely signed lobes are differentiated by color.}\label{fig:CoSbwf}
\end{figure}

\section{Appendix: Boltzmann transport from Wannier functions interpolation}

Prediction of electronic transport properties, using the Boltzmann transport equation (BTE), depends on the ability to accurately compute
and integrate band derivatives over the Brillouin zone. Usually this is achieved by fitting the electronic band to a smooth curve and
performing numerical derivatives, an approach that is sensitive to band crossings.
The Wannier representation of 		the electronic structure\cite{marzari_maximally_1997, yin_orbital_2006,volja_charge_2010, berlijn_can_2011}
  provides an optimized tight-binding model whose Hamiltonian can be directly differentiated to compute band velocities and effective
masses.\cite{marzari_maximally_1997,yates_spectral_2007}
An additional advantage of the approach is the possibility to separate the role of individual bands or band manifolds by
 projecting on minimal subspaces containing the most relevant degrees of freedom, using the disentanglement procedure.\cite{souza_maximally_2001}
For transport properties only a certain subset of the Bloch states near the Fermi level is relevant. In this work  we used
Maximally Localized Wannier Functions (MLWF) to
 derive the necessary ingredients for the BTE in the constant scattering time approximation. As a side product we obtained a description of
 the bonding states in terms of MLWF (Fig. \ref{fig:CoSbwf}).

Prototypical CoSb$_3$ is a semiconductor with two isolated valence manifold of 12 and 36 bands respectively and a conduction manifold that
 consists of infinite number
of entangled states above a LDA energy band gap of the order of $\approx$ 0.22  eV.
The lowest manifold of 12 valence states is mainly formed by Sb s-states.
The top 36-band manifold is constructed iteratively from the initial guess of the atomic Sb $s$- and $p$- and Co $d$-states.
Starting with a combination of on-site Co $d$-states and Sb $s$- or $p$-states we have converted original combinations of atomic orbitals to a
well localized set of Wannier functions with spreads in the range of 1.5-6.65  \AA$^2$.
Among all 48 valence states one can distinguish 12 Co states of t$_{2g}$ symmetry, 12 Sb-Sb bonding states and 24 Co-Sb bonding states
(Fig. \ref{fig:CoSbwf}).
To construct Wannier states for the conduction manifold for CoSb$_3$ we choose Bloch states in the energy range of 3.2 eV above the Fermi level.
MLWF states were obtained by iterative convergence starting with the initial guess of 24 gaussian-type orbital states
placed 1/4 of the Co-Sb bond length away from Co
 atoms along each of the 24 Co-Sb bonds. Using a similar approach we have also determined the basis of MLWFs for the PSTS systems.

Given the basis of MLWFs we can express matrix elements of the Hamiltonian in terms of Wannier functions.
     Matrix elements of a periodic operator {\it O} between Wannier states $n$ and $m$ are written as $O^W_{nm}(R)=\langle n0|O|mR \rangle$
The matrix element of the Hamiltonian at an arbitrary k point in the k-space (Ref 42.) can be obtained by inverse Fourier Transformation(FT) interpolation
$$H_{nm}({\bf k})=\sum_{{\bf R}} e^{i{\bf kR}}\langle n0|H|mR\rangle$$.
Due to the strong localization of MLWFs the Hamiltonian in the Wannier basis is sparse and one does not need the
original k-point mesh (used to construct the Wannier states) to be dense to obtain convergence for an arbitrary k point.
 For large systems, FFT scales much faster (O($N\log(N)$)) compared to
the scaling of the eigenvalue problem ( O($N^3$)).
The right hand side of the last equation can be differentiated analytically with respect to $k$ to obtain the matrix elements of the velocity operator.

$$v_{nm,\alpha}({\bf k})=\frac{\partial H_{nm}}{\partial k_{\alpha}}=\sum_{{\bf R}} e^{i{\bf kR}}(iR_{\alpha}) \langle n0|H|mR \rangle$$
As a last step, a rotation to the original set of the Bloch states is performed. This however, requires matrix multiplication of only very small matrices of M x M size, where M is the number of Wannier states.
As a result, this interpolation scheme is faster than the direct solution of the eigenvalue problem, but with the additional complexity of the initial Wannierization. It also resolves a number of difficulties associated with band crossings and avoided crossings which persist in traditional interpolation schemes.

\bibliographystyle{apsrev4-1}


%
 \end{document}